\documentclass[paper=a4,fontsize=11pt]{scrartcl} 

\usepackage[T1]{fontenc} 
\usepackage[sc]{mathpazo} 
\usepackage{palatino}

\usepackage{authblk}  

\usepackage{setspace} 
\usepackage{verbatim}
\usepackage[skip=4pt, indent=0pt]{parskip}
\usepackage{subcaption}
\usepackage{footmisc}
\makeatother

\renewcommand{\arraystretch}{1} 

\usepackage{sectsty} 
\usepackage[english]{babel} 

\usepackage{footmisc}
\usepackage[dvipsnames]{xcolor}

\usepackage{amsthm, stmaryrd}
\newtheorem{definition}{Definition}
\newcommand{\m}{\mathit}
\newcommand{\nm}{\m{nm}}
\newcommand{\relation}{R}

\newcommand{\Prolog}{\mathcal{P}}
\newcommand{\drule}{Q}

\DeclareOldFontCommand{\bf}{\normalfont\bfseries}{\mathbf}
\DeclareOldFontCommand{\it}{\normalfont\itshape}{\mathit}

\usepackage{listings}
\usepackage{geometry}
\usepackage{xcolor}
\usepackage{amsmath,amssymb,amsfonts,amsthm,mathtools} 

\usepackage{tikz,graphicx} 
\usetikzlibrary{shapes,automata,arrows,chains,matrix,positioning,scopes,calc,decorations.pathreplacing,fit,arrows.meta,decorations.pathreplacing}

\usepackage{multicol}
\usepackage{adjustbox}
\usepackage{longtable}
\usepackage{makecell}
\usepackage{rotating,lscape,pdflscape} 
\usepackage{float}
\usepackage{abstract}

\usepackage[labelfont=bf]{caption}

\renewcommand{\thefigure}{\arabic{figure}}
\renewcommand{\thetable}{\arabic{table}}

\usepackage{varwidth}
\usepackage{listings} 

\usepackage{tensor}
\usepackage{marvosym} 

\usepackage{bbding} 
\usepackage[flushleft]{threeparttable} 
\usepackage{booktabs} 
\usepackage{blkarray}
\usepackage{multirow}
\usepackage{rotating}
\usepackage{adjustbox}
\usepackage{bbm}
\usepackage{afterpage} 

\usepackage[round]{natbib} 

\usepackage[section]{placeins} 

\usepackage{color}
\definecolor{darkblue}{RGB}{0.0,0.0,80.0}
\definecolor{cranberry}{RGB}{193, 5, 52}
\definecolor{navy}{RGB}{26, 71, 111}

\usepackage{epigraph}

\usepackage{arydshln}

\usepackage[pdftitle={},pdfpagemode=UseOutlines,pdfborder={0 0 0},pdfauthor={},pdfsubject={},pdfkeywords={},pdfpagetransition=Dissolve,bookmarks=true]{hyperref}
\hypersetup{colorlinks,breaklinks,linkcolor=black,urlcolor=darkblue,anchorcolor=darkblue,citecolor=darkblue}
\usepackage{amsmath}
\usepackage{natbib} 
\usepackage[sectionbib]{bibunits} 
\defaultbibliographystyle{aer} 
\defaultbibliography{references}

\sectionfont{\fontsize{12}{12}\selectfont\bfseries}

\usepackage{chngcntr}  
\usepackage{apptools} 

\usepackage{tikz}
\usepackage{mathtools}
\usetikzlibrary{matrix,calc,positioning,shapes.geometric}
\begin{document}    

\sloppy 

\title{Large Language Models are overconfident and amplify human bias\thanks{We thank Chenrui Wang for excellent research assistance. We are grateful for comments from participants at seminars at the National University of Singapore, the University of Melbourne, Osaka University, Beijing Jiaotong University, the Stockholm School of Economics, the University of California - San Diego, the Herie School, and the University of Hong Kong. The first authorship is shared between Fengfei Sun and Ningke Li. Correspondence should be addressed to Kailong Wang (wangkl@hust.edu.cn) and Lorenz Goette (ecslfg@nus.edu.sg)}}  

\author[1]{Fengfei Sun}

\author[1]{Ningke Li}

\author[2]{Kailong Wang}

\author[1,3,4,]{Lorenz Goette}
\affil[1]{National University of Singapore,}
\affil[2]{Huazhong University of Science and Technology}
\affil[3]{Centre for Economic Policy Research}
\affil[4]{CESifo}

\date{}

\maketitle
\thispagestyle{empty}
\begin{abstract}
	
	Large language models (LLMs) are revolutionizing every aspect of society. They are increasingly used in problem-solving tasks to substitute human assessment and reasoning. LLMs are trained on what humans write and are thus exposed to human bias. We evaluate whether LLMs inherit one of the most widespread human biases: overconfidence. We algorithmically construct reasoning problems with known ground truths. We prompt LLMs to answer these problems and assess the confidence in their answers, closely following similar protocols in human experiments. We find that all five LLMs we study are overconfident: they  overestimate the probability that their answer is correct between 20\% and 60\%. Humans have accuracy similar to the more advanced LLMs, but far lower overconfidence. Although humans and LLMs are similarly biased in questions which they are certain they answered correctly, a key difference emerges between them: LLM bias increases sharply relative to humans if they become less sure that their answers are correct. We also show that LLM input has ambiguous effects on human decision making: LLM input leads to an increase in the accuracy, but it more than doubles the extent of overconfidence in the answers.

\end{abstract}

\clearpage

	Large language models (LLMs) are revolutionizing every aspect of society. Among other things, they are used in problem-solving tasks that require careful reasoning and assessment. While LLMs are well-known to reliably reproduce knowledge on which they have been trained \citep{zhao2023survey}, they are prone to making mistakes in reasoning tasks on which they have not been directly trained \citep{li2024drowzee, huang2023towards, chang2024survey,xu2023large,pawitan2025confidence} and their reasoning may collapse completely if problems become too complex \citep{shojaee2025illusion}.
    
    LLMs are trained on what humans write. This exposes them to human bias, and puts them at risk to develop them in their reasoning. One of the most prevalent biases in human judgment is overconfidence \citep{hoffrage2022overconfidence, kahneman2011thinking,moore2008trouble,malmendier2015verges}. Humans  are prone to underestimating the limits of their knowledge \citep{dunning2011dunning,kruger1999unskilled}, leading them to overestimate the quality of their judgments.
    \footnote{Various non-cognitive factors have also been shown to contribute to overconfidence, such as social signaling, \citep{burks2013overconfidence}, environmental factors \citep{goette2015stress}, or strategic considerations to feign confidence \citep{charness2018self} that may also provide an  evolutionary mechanisms for the trait \citep{johnson2011evolution,marshall2013evolutionary}. }

    Overconfidence has been shown to affect decision making in many areas.  In the professional domain, overconfident managers are more likely to use equity financing \citep{malmendier2005ceo, malmendier2011overconfidence} and more likely to engage in unprofitable mergers \citep{malmendier2008makes}, exhibiting little awareness or learning over time \citep{huffman2022persistent}. There is also evidence of overconfidence affecting everyday behaviors, such as working out regularly or dietary choices \citep{arni2020biased}, or financial investments \citep{barber2001boys}.  

    In this context, LLMs may hold, either, promise or peril. On the one hand, overconfident individuals may be challenged in their reasoning by LLM input, thus mitigating human bias. On the other hand, if overconfidence manages to enter LLMs, this may exacerbate bias in human decision makers. 
    
    \section*{The accuracy and overconfidence of LLMs}
    We first examine whether LLMs exhibit overconfidence in their own reasoning abilities. There are no general theoretical results to characterize in which contexts AI or LLM reasoning leads to biased answers. Conventional wisdom suggests the "bias in, bias out" hypothesis \citep{barocas2016big, mayson2019bias}: if models are trained on biased data, it is hypothesized that this biases its own responses in a similar way. However, this need not always hold. It is possible that training data exhibiting bias with regard to one attribute leads AI to be less biased towards the same attribute\citep{rambachan2020bias}
    \footnote{\citet{rambachan2020bias} show that in many realistic environments, bias in training data, stemming, e.g. from human bias against demographic group, can lead to less biased behavior of an AI trained on that data towards the group that was discriminated against in the training data. However, a corollary of their result is that biased training data will still affect other groups, with the AI being more biased towards them, even though humans don't display bias towards those groups.} 
    Inference and reasoning can further be affected by overconfidence: e.g., \citet{heidhues2018unrealistic} show that overconfident, but otherwise rational, agents may develop persistent bias instead of learning from feedback.  We thus aim to design an experiment to detect departures from unbiased calibration, and probe into the underlying mechanisms. We also designed our experiments such that they are sufficiently powered under the significance levels proposed by \citet{benjamin2018redefine} for novel findings. 
    
	To this end, we prompt five commonly used LLMs with 10,000 test cases to which we know the true answer. The questions are generated using the algorithm in \citet{li2024drowzee} to generate questions on which the LLMs are unlikely to be trained on. Thus, to arrive at an answer, the LLMs need to engage in reasoning on the spot. 
    \footnote{Several recent papers have examined measures of calibration that can be extracted from LLMs' intermediate outputs. Prior work has taken three main approaches to measuring LLM confidence: (1) logit-based estimation, which requires internal model access \citep{yin2023large,gpt4report2023,he2023investigating,zhang2024calibrating}; (2) direct confidence elicitation through prompting \citep{tian2023just,xiong2024can,wei2024measuring,wen2024mitigating}; and (3) auxiliary model approaches, ranging from single-model prediction \citep{ulmer2024calibrating}to multi-source integration methods \citep{zhao2024pareto}. These calibration studies predominantly rely on standard question-answering datasets, and it is likely that the LLMs have been trained on a large fraction of the questions used \citep{tian2023just,wei2024measuring,zhang2024calibrating,xiong2024can}. Thus, these studies do not address LLMs' confidence in questions that require them to reason. By contrast, we utilize algorithmically generated questions that guarantee minimal training contamination, allowing us to study confidence in answers that involve reasoning by LLMs.}
    We then ask the LLMs to assess the confidence in their answers, closely following similar protocols in human experiments. We elicit the LLM's confidence in the correctness of the answer, the facts used, and the reasoning. We set the LLM's temperature to zero, where possible.  
	
	We find that all five LLMs we examine are overconfident (Table \ref{tab:desc_stats}). The LLMs on average overestimate the probability that their answer is correct between 20\% (for GPT o1) and 60\% (for GPT 3.5), with the other models falling somewhere in between ($p<0.005$ in all cases). We find that more advanced models such as GPT 4o or GPT o1 have higher accuracy than GPT 3.5 or Llama 3.2, in line with earlier findings \citep{shahriar2024putting, zhong2024evaluation, dubey2024llama}. Interestingly, the confidence judgments across models are very similar, and in no way reflect the large differences in accuracy rates.   
    
    Next, we examine how, within each LLM model, accuracy is related to confidence across the different questions. Panels A and B of Figure \ref{fig:confidence_analysis} show the accuracy rates as a function of confidence, for the GPT and Llama models, respectively. The panel shows a strong and positive association between accuracy and confidence. The graphs also show that even when a model is fully confident in its answer, there is still substantial bias: Table \ref{tab:reg_acc_conf_10000} shows that the bias varies between 15\% and 21\% for the top performing models (GPT 4o, o1, and Llama 3.1), and even more for GPT 3.5 and Llama 3.2.  
    
    For all models except Llama 3.2, bias increases with lower confidence levels. The estimates in Table \ref{tab:reg_acc_conf_10000} quantify this effect: a 10\% drop in confidence is associated with a drop in accuracy of around 25\% for GPT 4o and GPT o1, and around 13\% for Llama 3.1 (see Table \ref{tab:reg_acc_conf_10000}). This implies that the bias gets larger as the models' confidence declines. 
    
    The strong association between confidence and accuracy also extends each component of the reasoning process in which the LLMs have to engage. For each question, we separately elicit the the LLM's confidence that the facts used are correct, and that the reasoning applied is correct. These confidence measures are highly correlated with confidence that the overall answer is correct (correlations range from 0.49 to 0.92 across models , see Table \ref{tab:corr_gpt_models} and Table \ref{tab:corr_llama_models} in the \emph{SI}). We also calculate the semantic similarity between the facts used and the reasoning applied with the sufficient set of facts, and confidence that the reasoning is correct (see section \ref{subsec:similarity_measurement} in the \emph{SI} for details). We find significant associations between confidence and similarity for, both, facts and reasoning ($p<0.001$ in all cases, see Tables \ref{tab:corr_gpt_models} and \ref{tab:corr_llama_models} in the \emph{SI}). Thus, there is a strong, but biased, relationship between accuracy and confidence in the answers, and this association extends to the necessary components for the reasoning. 
    
    We also examined the robustness of these results. In a first robustness test, we checked whether committing the model to a response and then asking about the confidence contributes to overconfidence. We set up analogous prompts where, instead of eliciting a direct yes/no answer, and subsequently a confidence statement, we ask directly "what is the probability that the answer is `yes'?" We then derive the implied response and confidence from those statements. We also elicit the same measure for the probability of the correct answer being "no." Table \ref{tab:prompt frames} in the \emph{SI} shows that there are no meaningful improvements in accuracy or calibration in either variants of the prompts or any LLM. In several cases, accuracy decreased and bias increased substantially in the alternative prompts. Thus, asking the LLM to first commit to an answer, and then assess the confidence in the answer does not contribute importantly to our baseline findings. Second, we varied the temperature of the LLM, i.e. the extent to which it picks the most likely response (at temperature zero, our baseline temperature) or allows for randomness in the response (at higher temperatures). As Table \ref{tab:summary_temp_performance} in the \emph{SI} shows, there is no quantitatively important effect of the temperature on accuracy or confidence: bias tends to decrease slightly at higher temperatures, by only in the order of a a few percentage points.\footnote{We also increased the temperature to 1.5, a value that would be considered too high for most applications. Indeed, most models were unable to produce complete answers at this temperature. The exeception is GPT 3.5, which still manages to produce answers, and enjoys higher accuracy. However, with a baseline accuracy of 35\%, even flipping a coin would increase baseline accuracy.} We also examine whether varying the temperature and prompt together produced more accurate and better calibrated results, but find little evidence thereof (see section \ref{sec:check temperature} in the \emph{SI} for details). 
    
    Finally, we also produced five replications of the results for GPT 3.5 and 4o at temperature 0, spanning over several months. We retrieve nearly identical results with regard to the answer and the confidence. The answers change in less than 2 percent in each iteration. The changes we observe also do not have a clear direction: overall, accuracy only increase by 1.7 percent after five replications (see section \ref{subsec:replication_results} in the \emph{SI}). 
    

    \section*{LLM vs. human overconfidence}
    We compare the accuracy, confidence and confidence gradient of LLMs to a human benchmark. In a large-scale online experiment on Prolific, we recruit participants to answer 10 questions of a subset of 2000 (see \emph{SI} section \ref{sec:human_benchmark}). Like the LLMs, the human participants have to pick one of the answers (Yes or No), and indicate the probability with which they believe that their answer is correct. Human participants answer 66\% of the questions correctly. This puts them on par in terms of accuracy with GPT 4o or Llama 3.1, but slightly behind GPT o1. Their confidence is significantly lower than that of any of the LLMs: human participants think their accuracy is 70\%, thus overestimating it by only 4\% on average ($p<0.001$, see Table \ref{tab:summary_human_confidence}  in the \emph{SI}). This is far lower overconfidence than any of the LLMs with comparable accuracy. There is a maked difference in shape of the distribution of confidence (see Figure \ref{fig:hist_conf_answer} in the \emph{SI}): compared to LLMs, humans' confidence appears far more nuanced, and with a mode (nearly 35\% of the answers) stating that it is 50:50 whether their answers are correct. 
    
    Panel C of Figure \ref{fig:confidence_analysis} shows the relationship between accuracy and confidence in the human sample, compared to the three most accurate LLMs. Two striking features emerge from the graph: first, when humans are 100\% sure that their answer is correct, their accuracy rate is significantly lower at 81\% ($p<0.001$, see column 1 in Table \ref{tab:sub_reg_conf_acc} ). Thus, for questions that humans consider relatively easy and are thus 100\% sure of their answer, they are highly overconfident. This level of overconfidence is roughly on par with the same conditional accuracy rates of GPT 4o or Llama 3.1. However, while originating from the same level of bias when 100\% sure of their answers, the confidence gradients between LLMs and humans are sharply different. As discussed before, LLMs become more biased as they enter more uncertain territory. In sharp contrast, the human confidence gradient is only about 0.5, indicating that human bias decreases as humans become less certain. As can be seen in the figure, humans eventually end up slightly underconfident: when participants think their accuracy is only 50\%, it is actually slightly higher (54\%, significantly higher than they expect $p<0.01$). 

    We interpret this sharp difference in the confidence gradient as a stark manifestation of the Dunning-Kruger effect \citep{dunning2011dunning,kruger1999unskilled}. Our test cases require LLMs to reason to answer them, as they are unlikely to be trained on the answer. This may cause LLMs being "trapped" in their prediction model in a way that is hard for them to overcome: the model-internal mechanisms cannot provide them with a sense of what knowledge is not part of their training data. The LLMs pick the answer that is most consistent with what they were trained on, and form an estimate of their accuracy based on what is in their training data. As LLMs move into less certain territory, this problem gets stronger, and causes LLMs to have much less of a sense of the limits of their knowledge. Human participants may recognize in a question, e.g., that one of the names is a Roman emperor, and realize that she does not know much about Roman history. By contrast, in an LLM, there is no mechanism to tell it that there might be other Roman public figures that it wasn't trained on. This results in overconfidence far stronger than what is observed in humans though the mechanisms of the Dunning-Kruger effect. Interestingly, while more advanced models like GPT 4o and GPT o1 have higher accuracy rates for test cases where they are certain their answer is correct, they also display a significantly stronger confidence gradient, and, hence a stronger Dunning-Kruger effect.

    \section*{Exposure to LLM input}
    While the computer science literature has documented limitations in LLMs' ability to reason, human users may not be aware what type of questions LLMs have been directly trained on, and which questions involve reasoning by LLMs. In a second experiment, we test how human performance, in terms of, both, accuracy and bias, is affected by exposure to LLM answers and LLMs' confidence in their answers. We test this in an information-intervention design: like in the previous experiment, participants answer questions and state their estimate of the probability of their answer being correct. In a second stage, participants in the "LLM Answer" condition are then provided with the answer that the LLM picked for this question. We randomly choose to show the answer from, either GPT 4o, o1, or Llama 3.1, i.e. only the three models with the highest accuracy. Participants are then given the opportunity to revise their own answer and confidence in the answer. In the "LLM Answer + Confidence" condition, participants are shown the LLM's answer as well as the LLM's confidence in its answer. In a control condition, participants are simply given the opportunity to revise their answer (see \emph{SI} section \ref{sec:llm_exposure_experiment} for more details). 
    
    Figure \ref{fig:treatment_effects} shows how exposure to LLM input changes the accuracy (Panel A) and bias with regard to accuracy (Panel B). Exposure to, either, LLM answer or answer plus confidence increases the accuracy of the human participants' answers between 5.6 and 7 percentage points. However, both conditions the participants' confidence in their answers by even more, increasing bias by 4.2 percentage points in the answer condition, and 7.6 percentage points in the Answer + Confidence condition. Displaying the LLM's confidence does not lead to a significant increase in accuracy, but  increases bias significantly more than only providing the LLM answer ($p<0.01$, see \emph{SI} Table \ref{tab:004_1_treatment_effects}). The amount of bias that exposure injects into humans' assessments is large: at baseline, participants exhibited only moderate overconfidence, of around 4\%. Thus, exposure to LLM input doubled (in LLM Answer), or nearly tripled (in LMM Answer + Confidence) their bias.

    We next examine which types of users of LLMs are driving these overall treatment effects. LLMs help improve the accuracy of individuals with low baseline confidence in their answer, raising it by 8.6 and 11.9 percentage points, respectively (see Panel A of Table \ref{tab:heter_treatment_effect}. However, LLM exposure also substantially increases bias: human confidence in the revised answer increases by far more than what is justified by the increase in accuracy. Bias increases by 7.0 percentage points in the LLM Answer condition, and by 14.1 percentage points when, in addition, the LLM confidence is also displayed. The LLM answer nearly triples bias, and displaying the LLM confidence more than quadruples it in questions where individuals have below-median confidence at baseline. Interestingly, displaying LLM confidence is not necessary to increase bias in humans. By contrast, individuals with above-median confidence in their answer experience seem `immovable:' they experience neither a gain in accuracy nor a increase or decrease in bias. 

    We also explore how heterogeneity in LLMs' confidence affect exposure to LLM input. Table \ref{tab:heter_treatment_effect} shows the results.  Panel B shows how low-confidence (80\% to 90\%) and high-confidence (90\% - 100\%) answers from LLMs affect the users accuracy and bias. Low-confidence answers from the LLM do not increase accuracy (the point estimate is slightly negative) in either condition. However, bias substantially increases. By contrast, answers in which LLMs have more than 90\% confidence increase accuracy, but fail to decrease bias.

    \section*{Discussion}

    
    Overall, our results suggest that LLMs are far more overconfident than humans in tasks that involve LLM reasoning. Humans, in turn, do not sufficiently discount LLM advice. On average, exposure to LLMs input leads humans to become more accurate, but also more biased. The analysis in sub-sambples shows that this effect is driven by humans with low baseline confidence in their answer for, both, the accuracy and bias.

    While LLMs generate output that is highly valuable to users by tapping into answers to questions on which they are pre-trained, our analysis shows tasks that require LLMs to reason may not increase human welfare. We examine this formally in section \ref{sec:model} of the \emph{SI}. We set up a model in which payoffs depend on, both, the probability of getting the answer correct, and a costly investment that an individual makes. If the payoffs are largely proportional to the probability of giving a correct answer, then LLMs  increase welfare. In such a setting, bias per se does not have any costs. However, many tasks also require a decision or investment depending on the assessed probability of success. In these environments, overconfidence leads to over-investment, and reduces expected payoffs relative to what would be optimal. It is easily possible that the increase in bias offsets the gain in accuracy (see Result 3 in \emph{SI} section \ref{sec:model}). Quite intuitively, the more elastic investment responds to the perceived probability of success, the greater the scope for increased bias from LLM exposure to reduce welfare. The model also makes more subtle predictions. Suppose, as our data suggest, that individuals are biased at baseline. In this case, LLM exposure can lead to lower welfare even if only the probability of providing a correct answer increases. The reason is that the higher accuracy increases investment. Because marginal costs of investment are increasing, this makes the over-investment from bias more costly. As we show, this effect can be so strong that it offsets the gains from higher accuracy (see Result 2 in \emph{SI} section \ref{sec:model}).  This highlights an important warning: even intuitive indicators such as increased accuracy from LLM exposure may be not be indicative of higher payoffs. The opposite may, in fact, be the case, as our examples demonstrate.

    In assessing the benefits from LLMs, is also noteworthy that LLMs and human participants express uncertainty about different questions. In our sample, the correlation between average human confidence and average LLM confidence is only 0.082, see \emph{SI} Table \ref{tab:subsample_correlation_matrix}. This tends to improve welfare, as individuals who are uncertain about an answer receive roughly average-quality LLM answers, which have a higher accuracy than humans who are uncertain at baseline. If uncertainty were more correlated, LLM input would lead to lower accuracy for low human baseline confidence, but bias would increase substantially (see \emph{SI}, Table \ref{tab:009_subsample_treatment_effects}.) Correspondingly, high-confidence LLM answers would be `wasted' on high-confidence humans, where little improvement in accuracy is possible thus confirming their answers and further increase their bias.


    The implications of LLM overconfidence span practical, ethical and technical dimensions across stakeholder groups. Our results provide users with a new benchmark for how users of LLMs should approach reasoning from LLMs, counteracting the widespread overestimation of LLM capabilities \citep{KLINGBEIL2024108352, holbrook2024, info:doi/10.2196/56764}.  
    Secondly, our results provide guidance for the future development of LLMs. Architectural choices in training objectives currently prioritize fluency over accuracy, incentivizing fabricated but coherent responses \citep{yin2023large, fi15100336}.  The lack of a built-in uncertainty correction mechanism generates hallucinated responses and overconfidence \citep{shorinwa2024survey}. Current evaluation metrics~\citep{wei2024measuring, geng-etal-2024-survey} fail to capture calibration quality, necessitating new frameworks for quantifying and gauging overconfidence such as the one we presented here. Developing novel interpretability frameworks~\citep{wen2024from, 9857619} that pinpoint how and why overconfident and biased predictions emerge can support clearer communication of model uncertainty and limitations to users.     
    
    Our results indicate that making LLMs more knowledgeable (in terms of training data, or parameters in the model) is unlikely to remove overconfidence: while we find that overconfidence decreases with knowledge when LLMs are certain to be correct, they also show that the Dunning-Kruger effect is significantly stronger for larger LLMs as uncertainty increases. In other research, we examine the prevalence of framing effects in LLM answers \cite{fu2025llms}. We find strong framing effects in all models. Larger and newer models display weaker framing effects, but the gains appear to level off: for instance, we find no significant reduction in framing effects between OpenAI's o1 and and the most recently released o3. This points to the importance of incorporating validation mechanisms to check the LLM's reasoning process \citep[see, e.g.][]{binder2024looking}, rather than simply trying to make models larger. This line of research has the potential to stimulate new algorithmic approaches and training strategies, ultimately fostering accountable, trustworthy, and ethically sound AI systems~\citep{NEURIPS2022_ff887781, li2024overconfident}.

    More broadly, this research highlights the usefulness and importance of evaluating the presence, extent and consequences of behavioral biases in LLM reasoning. There are many other potential biases well-known in behavioral science, such as basic violations of rationality \citep{chen2023emergence}, framing effects \citep{tversky1981framing,kahneman1984choices}, gender bias \citep{sun2019mitigating}, the conjunction fallacy \citep{tversky1983extensional,busemeyer2011quantum}, or conformation bias \citep{klayman1995varieties,nickerson1998confirmation} that permeate training data generated by humans. Behavioral science has established paradigms to measure these departures from rationality. A nascent literature examines how individuals respond to (potentially biased) AI input.  Fruitful collaboration with computer science awaits to explore the overarching mechanisms of how they manifest in LLMs, and how they can be reduced.




\clearpage 

\section*{Tables and figures}

\begin{table}[h!]
\centering
\begin{threeparttable}
	\begin{tabular}{lccc}
		\toprule \addlinespace
		\textbf{Model} & \textbf{a) Accuracy} & \textbf{b) LLM confidence} & \textbf{c) Bias} \\
		 & \textbf{rate}& \textbf{in answer} &  \\ \addlinespace
		\midrule
		GPT 3.5 & 0.35 & 0.94 & 0.59 \\
		&       & (0.045) & (0.473) \\ \addlinespace
		GPT 4o  & 0.63 & 0.94 & 0.30 \\
		&       & (0.053) & (0.470) \\ \addlinespace
		GPT o1  & 0.73 & 0.95 & 0.22 \\
		&       & (0.044) & (0.433) \\ \addlinespace
		Llama 3.1 & 0.63 & 0.86 & 0.23 \\
		8B&       & (0.102) & (0.466) \\ \addlinespace
		Llama 3.2 & 0.61 & 0.94 & 0.33 \\
		3B&       & (0.075) & (0.486) \\  \addlinespace
		\bottomrule
	\end{tabular}
	
\caption{\textbf{Accuracy, confidence, and bias across models}} \label{tab:desc_stats} \medskip 
\begin{tablenotes}
\small 
\item \textit{Notes:} $N = 10,000$ questions submitted to LLM (see \ref{tab:reg_acc_conf_10000} for exact number of observations in each LLM). Accuracy rate is the fraction of correct answers. Confidence is elecited from LLM: "What is the probability that your answer is correct? Bias is defined as confidence - accuracy rate. Standard deviations in parentheses (divide by 100 to obtain approximate SE of means). All accuracy rates and confidence measures are significantly different between models. Average bias in each model is significantly different from zero.  ($p<0.001$, see \emph{SI} Table \ref{tab:overconfidence_test}.)
\end{tablenotes}
\end{threeparttable}

\end{table}

\begin{table}[h!]
\centering
\begin{threeparttable}
    \begin{tabular}{lccccc}
\toprule \addlinespace
& GPT 3.5 & GPT 4o & GPT o1 & Llama 3.1 & Llama 3.2 \\
\addlinespace \midrule \addlinespace
Confidence & 1.31 & 2.44 & 2.48 & 1.33 & 0.56 \\
gradient & (0.10) & (0.09) & (0.15) & (0.05) & (0.07) \\ \addlinespace
Predicted accuracy & 0.42 & 0.79 & 0.85 & 0.81 & 0.65 \\
if fully confident & (0.01) & (0.01) & (0.01) & (0.01) & (0.01) \\
\addlinespace \midrule \addlinespace
$N$ & 9,795 & 9,973 & 9,990 & 9,750 & 9,910 \\
\addlinespace \bottomrule
\end{tabular}
\caption{\textbf{The confidence gradient in accuracy}}
\label{tab:reg_acc_conf_10000}
\medskip
\begin{tablenotes}
\small 
\item 
\textit{Notes:} Dependent variable is correct answer (= 1). The table presents OLS estimates of the gradient of accuracy with respect to self-reported confidence in answer across models (see Table \ref{tab:desc_stats} for definition). Confidence levels are normalized by subtracting 1 from the original confidence scores. Robust standard errors in parentheses. All predicted accuracy rates at full confidence are different from 1 at $p<0.005$. Confidence gradients are all greater than one ($p<0.005$), except for Llama 3.2. \vspace{1cm}
\end{tablenotes}
\end{threeparttable}
\end{table}

\begin{table}[h!]
\centering
\begin{threeparttable}
\begin{tabular}{lcccc}
\toprule \addlinespace
Treatment interacted with:& \multicolumn{2}{c}{Human Baseline Confidence} & \multicolumn{2}{c}{LLM Confidence} \\
\cmidrule(lr){2-3} \cmidrule(lr){4-5}
Dependent variable:& $\Delta$ Accuracy & $\Delta$ Bias & $\Delta$ Accuracy & $\Delta$ Bias \\ \addlinespace
\midrule \addlinespace
LLM Answer & 0.086 & 0.070 & -0.033 & 0.133 \\
& (0.015) & (0.017) & (0.018) & (0.020) \\
\addlinespace
LLM Answer + Conf. & 0.119 & 0.141 & -0.028 & 0.160 \\
& (0.017) & (0.018) & (0.020) & (0.021) \\
\addlinespace
LLM Answer & -0.064 & -0.063 & 0.136 & -0.138 \\
$\times$ High Confidence & (0.018) & (0.019) & (0.020) & (0.021) \\
\addlinespace
LLM Answer + Conf. & -0.091 & -0.124 & 0.153 & -0.119 \\
 $\times$ High Confidence & (0.019) & (0.019) & (0.024) & (0.024) \\
\midrule
Observations & 11,610 & 11,610 & 10,957 & 10,957 \\
\bottomrule
\end{tabular}
\caption{\textbf{Subsample Analysis of LLM Exposure Experiment}}
\label{tab:heter_treatment_effect}
\medskip
\begin{tablenotes}
\small 
\item \textit{Notes:} OLS estimates, displaying the difference-in-difference estimates of LLM exposure by different subgroups. Standard errors in parentheses are adjusted for clustering at the participant and question level. $N = 1,161$ subjects answered 10 questions each in a intervention study design (see section \ref{sec:llm_exposure_experiment} in the \emph{SI} for details of the design). 
High Confidence is defined as confidence above 0.9 for LLM and above the median for human baseline. The regressions contain a constant term (not shown), and also control for high confidence indicators, though these results are not displayed in the table. 
Figure \ref{fig:heter_treatment_effects} visualizes the overall effects in the different subgroups. All main treatment effects are significantly different from control condition ($p<0.005$ in all cases). All treatments have differential impact on high vs. low confidence subgroups ($p<0.005$ in all cases). 
\end{tablenotes}
\end{threeparttable}
\end{table}

\clearpage

\begin{figure}[h!]
    \centering
    \includegraphics[width=\textwidth]{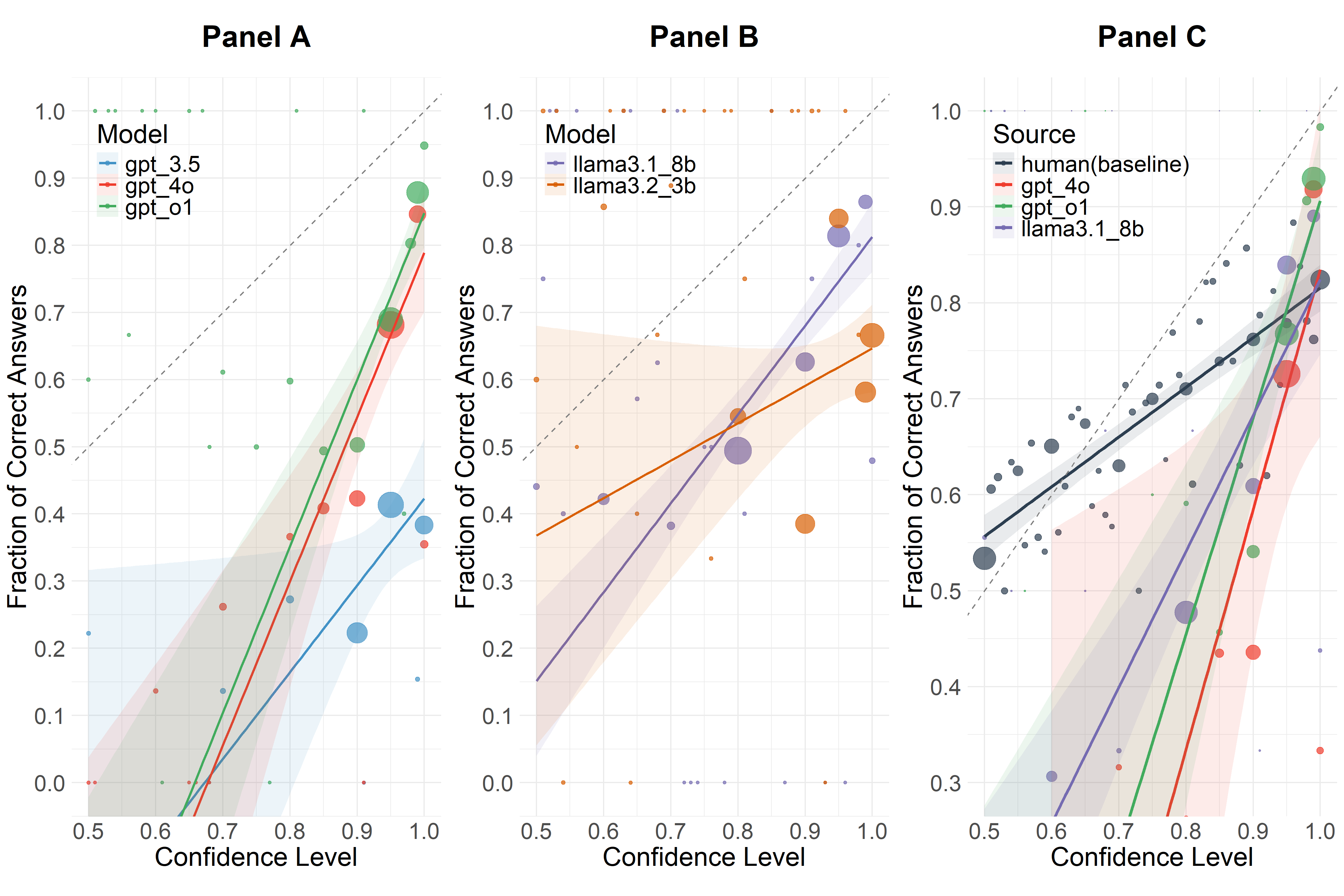}
    \vspace{-0.2cm}
    \caption{\textbf{LLM Confidence and Accuracy}}
    \label{fig:confidence_analysis}
    \medskip
    \begin{minipage}{0.95\linewidth}
        \small
        \emph{Notes:} Panel A shows the relationship between accuracy and confidence for GPT models, and Panel B shows the same relationship for Llama models. Panel C compares the accuracy-confidence relationship between advanced LLM models and human responses on the question subset for the human baseline benchmark (see section \ref{sec:human_benchmark} of the \emph{SI} for details on the design). The size of circles in each panel scales with the number of observations, and all slopes are significant ($p < 0.001$ in all cases; see Table~\ref{tab:reg_acc_conf_10000} for Panel A and B, and Table~\ref{tab:sub_reg_conf_acc} in Supplementary Information for Panel C). Shaded areas indicate 95\% confidence bands.
    \end{minipage}
\end{figure}

\clearpage 

\begin{figure}
    \centering
    \includegraphics[width=\linewidth]{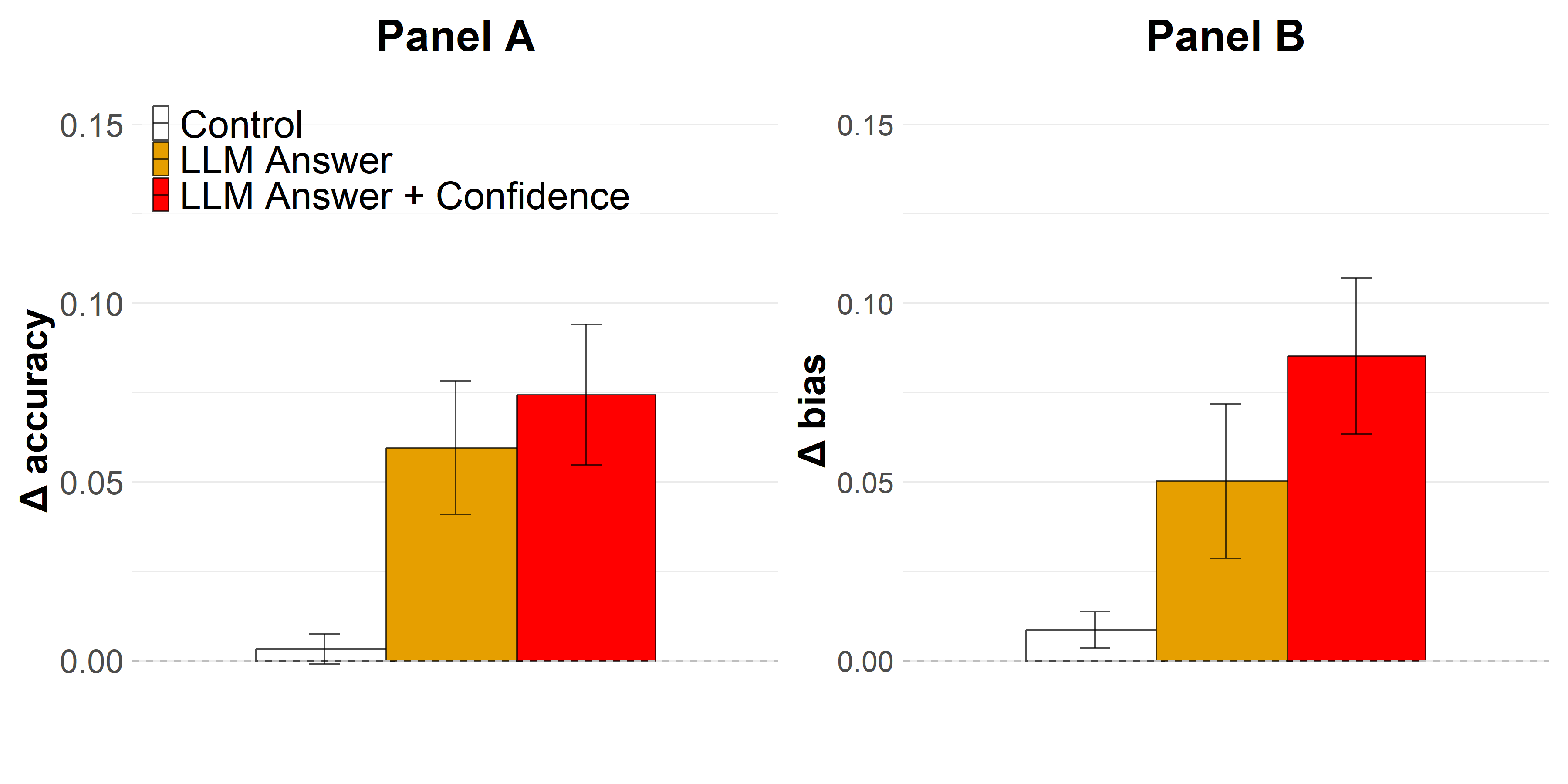}
    \vspace{-0.2cm} 
    \caption{\textbf{Treatment Effects on Changes in Accuracy and Bias}}
    \label{fig:treatment_effects}
    \medskip
    \begin{minipage}{0.95\linewidth}  
        \small
        \emph{Notes:} Panel A shows the treatment effects on change in accuracy across experimental conditions. Panel B shows the treatment effects on the change in bias across the same conditions. Error bars represent 95\% confidence intervals. Both treatment conditions have significant effects on accuracy and bias relative to the control group ($p<0.005$ in all cases). LLM Answer + Confidence has a significantly larger effect on on bias than LLM Answer $(p<0.005)$. For more details, see \emph{SI} Table \ref{tab:004_1_treatment_effects}. 
    \end{minipage}
\end{figure}

    \clearpage
\bibliographystyle{aer}  
\bibliography{library_latex}

\clearpage
\appendix
\renewcommand{\thefigure}{S\arabic{figure}}
\renewcommand{\thetable}{S\arabic{table}}
\setcounter{figure}{0}
\setcounter{table}{0}

\appendix
\tableofcontents

\clearpage

\section{The LLM Experiment}
\subsection{Question Generation}

\subsubsection{Triple Extraction from Wikipedia}
The first step involves extracting fundamental facts from the input knowledge data into structured triples for logical reasoning. Wikidata is a structured knowledge base that provides a vast collection of interconnected data on Wikipedia, including entities and their relationships, making it an ideal resource for extracting structured triples. We select the top-ten popular categories from Wikidata' entity and relation classifications to guide the extraction of facts (for category details, see Table \ref{tab:question_features}). Each fact is formatted as a triple (subject, predicate, object), where the subject and object represent entities, and the predicate describes their relationship. The extraction is done by searching through the knowledge base, retrieving all facts related to each entity and its corresponding relations. This methodical organization of triples establishes a solid foundation for subsequent logical reasoning operations.

\subsubsection{Logical Reasoning Process}
\paragraph{Step 1: Rule Generation. } 
In this step, an automatic rule generator is designed to iterate its predicates and generate the derivation rules $\drule$ according to the relation category. 
There are five reasoning types of our interest: Negation Reasoning, Symmetric Reasoning, Inverse Reasoning, Transitive Reasoning, and Composite Reasoning. The definition and organization of facts using each reasoning type is detailed in the following paragraphs.

\noindent\textbf{\textit{Rule\#1: Negation Reasoning.}} Based on a given factual knowledge, we can determine whether the opposite of this fact is correct or incorrect by applying Definition~\ref{def:neg}.
\begin{definition}\label{def:neg}

\textbf{Negation Reasoning Rule $[Neg]$.} Given a factual knowledge triple $(s, \nm, o)$, then we can derive the new triple $(s, \overline{\nm}, o)$ is not valid. $\overline{\nm}$ indicates the negation of the relation $\nm$.
\end{definition}

\[
    \frac{
        \begin{array}{c}
          \nm(s, o)\\
        \end{array}
    }{
    \begin{array}{c}
     \neg ~ \overline{\nm}(s, o)   
      \end{array} 
      }
    [Neg]
\]
An example of this type of rule is: $\frac{
        \begin{array}{c}
          \m{was}(s, o)\\
        \end{array}
    }{
    \begin{array}{c}
     \m{\neg ~ wasn't}(s, o)   
      \end{array} 
      }
    [Neg]$.
    
With this rule, from the triple \emph{(Haruki Murakami, did not win, the Nobel Prize in Literature in 2016)}, we derive that the negation of this triple \emph{(Haruki Murakami, won, the Nobel Prize in Literature in 2016)} contains false factual knowledge. 

\noindent\textbf{\textit{Rule\#2: Symmetric Reasoning.}} In symmetric relations, if the subject and object in a triple maintain coherence upon interchange, a new triple can be deduced in accordance with Definition~\ref{def:symmetric}.
\begin{definition}\label{def:symmetric}
\textbf{Symmetric Reasoning Rule $[Sym]$.} Given a factual knowledge triple $(s, \nm, o)$, then we can derive a new triple $(o, \nm, s)$.
\end{definition}

\[
    \frac{
        \begin{array}{c}
          \nm(s, o)\\
        \end{array}
    }{
    \begin{array}{c}
      \nm(o, s)   
      \end{array} 
      }
    [Sym]
\]

An example of this type of rule is: $\frac{
        \begin{array}{c}
          \m{different\_from}(s, o)\\
        \end{array}
    }{
    \begin{array}{c}
      \m{different\_from}(o, s)   
      \end{array} 
      }
    [Sym]$.

With this rule, from the original triple \emph{(Haruki Murakami, different\_from, Haruki Uemura)}, we derive a new triple \emph{(Haruki Uemura, different\_from, Haruki Murakami)} (Haruki Uemura is a Japanese judoka). Note that the symmetric reasoning rule is primarily utilized within the composition reasoning rule~(to be detailed next) and does not introduce new knowledge on its own.

\noindent\textbf{\textit{Rule\#3: Inverse Reasoning.}} In an inverse relation, the subject and object can be reversely linked through a variant of the original relation, as defined in Definition~\ref{def:inverse}.
\begin{definition}\label{def:inverse}
\textbf{Inverse Reasoning Rule $[Inverse]$.} Given a factual knowledge triple $(s, \nm, o)$ and a reversed relation $\nm'$ of $R$, then we can derive a new triple $(o, \nm', s)$.
\end{definition}

\[
    \frac{
        \begin{array}{c}
          \nm(s, o), \nm'=Reverse(\nm)\\
        \end{array}
    }{
    \begin{array}{c}
      \nm'(o, s)   
      \end{array} 
      }
    [Inverse]
\]

An example of this type of rule is: $\frac{
        \begin{array}{c}
          \m{influence\_by}(s, o)
        \end{array}
    }{
    \begin{array}{c}
      \m{influence}(o, s)   
      \end{array} 
      }
    [Inverse]
    $.
With this rule, from the triple \emph{(Haruki Murakami, influence\_by, Richard Brautigan)}, we can derive a new triple \emph{(Richard Brautigan, influence, Haruki Murakami)}.

\noindent\textbf{\textit{Rule\#4: Transitive Reasoning.}} In transitive relations, if the object in one triple is the subject of the second triple, we can therefore derive a new triple following the Definition~\ref{def:transitive}.
\begin{definition}\label{def:transitive}
\textbf{Transitive Reasoning Rules $[Trans]$.} Given two factual knowledge triples $(s_1, \nm, o_1)$ and $(s_2, \nm, o_2)$, if $o_1$ is semantically equivalent to $s_2$, then we can derive a new triple $(s_1, \nm, o_2)$.
\end{definition}

\[
    \frac{
        \begin{array}{c}
          \nm(s_1, o_1),~ \nm(s_2, o_2), ~o_1 = s_2\\
        \end{array}
    }{
    \begin{array}{c}
      \nm(s_1, o_2)   
      \end{array} 
      }
    [Trans]
\]
An example here is: 
\[
    \frac{
        \begin{array}{c}
          loc\_in(s_1, o_1),~ loc\_in(s_2, o_2), ~o_1 = s_2\\
        \end{array}
    }{
    \begin{array}{c}
      loc\_in(s_1, o_2)   
      \end{array} 
      }
    [Trans].
\]
With this rule, from triples \emph{(Haruki Murakami, locate\_in, Kyoto)} and \emph{(Kyoto, locate\_in, Japan)}, we derive a new triple \emph{(Haruki Murakami, locate\_in, Japan)}.

\noindent\textbf{\textit{Rule\#5: Composite Reasoning.}} The previous four reasoning rules are all meta-rules capturing the most basic and fundamental logical relations among the facts and rules. 
Several basic reasoning rules can be chained together to form a composition reasoning rule if the relations in the rules are logically related.
Composite reasoning rules can generate knowledge that requires multiple steps of reasoning.

\begin{definition}\label{def:composite}
\textbf{Composite Reasoning Rules $[Comp]$.} Given multiple basic reasoning rules or predicates $[Rule_i] \in \{[Neg],[Sym],[Inverse],[Trans],[Predicates]\}$, we can chain them up to form a new composite reasoning rule.
\end{definition}

\[ 
   \frac{
    \frac{
        \frac{
        \begin{array}{c}
         \nm_{1\_Rule_1}(...), \nm_{2\_Rule_1}(...), ...\\
          \end{array} 
        }{
        \begin{array}{c}
          R_1
          \end{array} 
          }
        [Rule_1],\; ...
    }
    {
    \frac{\frac{
        \begin{array}{c}
          ...
          \end{array} 
        }{
        \begin{array}{c}
          ...
          \end{array} 
          }
        [...],\;...
    }{
        \frac{
        \begin{array}{c}
          \nm_{1\_Rule_i}(...), \nm_{2\_Rule_i}(...), ...\\
          \end{array} 
        }{
        \begin{array}{c}
         R_{\m{i}} 
          \end{array} 
          }
        [Rule_i], \;  ...
    }
   }}{
        \begin{array}{c}
          R_{\m{new}} 
          \end{array} 
   }
    [Comp]
\]

\paragraph{Step 2: Prolog Inference.} 
With the predetermined rules, we can be assisted with the Prolog engine, asserting all the related triples and consulting the reasoning rules. We use $\llbracket \relation \rrbracket_{\Prolog}$ to denote the query results of $\relation$ w.r.t the Prolog program $\Prolog$. 
When $\relation$ contains no variables,
it returns Boolean results indicating the presence of the fact; otherwise, it outputs all the possible instantiations of the variables. 
Then by obtaining solutions from Prolog queries, we can generate new knowledge triples based on the entities and their relations provided. For each instantiation that contains one subject ``s'' and one object ``o'', we then compose them with the new predicate to form the newly derived triple.  
They are later used as the complex knowledge to generate test cases and the corresponding oracles.

\subsubsection{Predicate Verification Process}
From the new knowledge triples, we manually review the predicate component to ensure the accuracy and validity later in the generated question-answer pairs. After removing potentially ambiguous knowledge, we ultimately retain 476 predicates along with their corresponding triples.

\subsubsection{Question Formulation and Refinement}
After generating triples from the Wikipedia database and manually verifying predicate verbs, we utilize GPT-4o to automatically create corresponding initial questions. To enhance the naturalness of these questions and minimize potential misinterpretations by generative AI in subsequent steps, we employed a sophisticated prompt for question refinement. This prompt was designed to check for grammatical errors and improve overall question formulation. The full prompt is as follows:

\begin{quotation}
Please formulate a question of the form ``Is it true that ...'', using the following structure: the subject is ``[subject]'', the object is ``[object\_]'', and the predicate in short-hand notation is ``[predicate]''. Before generating the final question, please perform the following checks and adjustments:

\begin{enumerate}
    \item Predicate Validity Check: Examine whether the given predicate ``[predicate]'' makes grammatical sense in English. If it doesn't, please reformulate it to ensure it follows proper English syntax and conventions. For example, \texttt{same\_instance\_of} might be reformulated to ``is the same instance as'' or ``is identical to''.
    
    \item Sentence Completeness Check: Assess whether the current predicate is sufficient to generate a complete, coherent sentence. If not, extend it appropriately without altering its core meaning. For example, \texttt{same\_named\_after} could be extended to ``is named after the same person/thing as''.
\end{enumerate}

After making these adjustments, generate a grammatically correct and complete question that accurately represents the relationship between the subject and object using the (potentially modified) predicate. Your final output should be a single, well-formed question in the format ``Is it true that [subject] [predicate] [object]?'', where [predicate] has been checked and adjusted if necessary.
\end{quotation}

This prompt ensures that the generated questions are grammatically correct, contextually appropriate, and free from potential underscoring issues, thereby reducing the risk of misinterpretation in subsequent AI processing stages.
 Using this methodology, we created 10,000 questions with balanced categories and reasoning types for testing purposes.

\subsection{Response Generation and Confidence Measurement}
We employed APIs on representative closed-source LLMs (GPT-3.5, GPT-4o, and GPT-o1) and open-source LLMs (Llama 3.1 8b and Llama 3.2 3b) to answer these questions independently, with temperature set to 0. We developed a prompt that not only elicits answers but also directs the AI to evaluate its own confidence levels regarding answer correctness, reasoning process, and facts used. The prompt used for generating these confidence-aware responses is as follows:

\begin{quotation}
Please answer the following yes/no question '[question]'. This question has only one correct answer. Follow these steps:

\begin{enumerate}
    \item Think through the question step-by-step, employing a human-like reasoning process.
    
    \item Pick the answer that you think is correct and begin with:
    \begin{itemize}
        \item ``The answer to the question is yes. ... (reason)''
        \item ``The answer to the question is no. ... (reason)''
    \end{itemize}
    Even if you are unsure about the answer, pick the one that you think is more likely correct, and give your reasons.
    
    \item Explain your reasoning process in detail.
    
    \item List the key pieces of knowledge used in your reasoning, presented as declarative sentences and enumerated.
    
    \item After providing your answer, evaluate your response in three aspects:
    \begin{itemize}
        \item What is your estimate of the probability (in percent) that your answer is correct?
        \item What is your estimate of the probability (in percent) that the facts underlying your answer are correct?
        \item What is your estimate of the probability (in percent) that the reasoning underlying your answer is correct?
    \end{itemize}
\end{enumerate}
\end{quotation}

\subsection{Similarity Measurement}\label{subsec:similarity_measurement}
\subsubsection{Methodology Framework}

The similarity score calculation methodology compares two sets of triples for each question ID: evidence triples and response triples (extracted from the model's answer). The calculation follows a three-step process.

\subsubsection{Facts Similarity Algorithm}

\paragraph{Step 1: Subject Matching. } 
For each evidence triple, we identify relevant response triples based on subject matching criteria:
\begin{itemize}
   \item For subjects containing 1-2 words: Match is established if the response subject contains at least one identical word
   \item For subjects containing 3+ words: Match is established if the response subject contains at least two identical words
\end{itemize}

\paragraph{Step 2: Object Similarity Calculation.} 
After identifying relevant subjects, we calculate object similarities using the following process:

\paragraph{Case 1:} For evidence triples with shared subjects:
\begin{itemize}
   \item Given evidence triples $[(A,\_,B), (A,\_,C)]$ and response triples $[(A,\_,B'), (A,\_,C'), (A,\_,D)]$
   \item Calculate similarities: $sim(B,B')$, $sim(C,C')$
   \item For non-direct matches (e.g., object $D$), calculate $sim(D,B)$ and $sim(D,C)$. Keep the higher similarity score for the final calculation 
   \item Store all similarity scores in $similarity\_list$
\end{itemize}

\paragraph{Case 2:} For evidence triples with different subjects:
\begin{itemize}
   \item Given evidence triples $[(A,\_,B), (C,\_,D)]$ and response triples $[(A,\_,B'), (C,\_,E)]$
   \item Calculate and store similarities: $sim(B,B')$, $sim(D,E)$
\end{itemize}

\paragraph{Step 3: Final Score Calculation} 
We propose two scoring mechanisms to evaluate LLM's response quality from different perspectives:

\paragraph{(1) Average Similarity Approach} 
This approach considers all generated content from the LLM, including potentially irrelevant information:
\begin{itemize}
   \item Store all calculated similarity scores
   \item Final fact score = average of all similarity scores
\end{itemize}
This method provides a comprehensive evaluation of the LLM's output, including its tendency to generate extraneous information.

\paragraph{(2) Maximum Similarity Approach}
This approach focuses on the LLM's best attempts to match the evidence:
\begin{itemize}
   \item For each evidence triple, retain only the highest similarity score among all matching response triples
   \item Final fact score = average of these maximum similarity scores
\end{itemize}
This method evaluates the LLM's ability to generate the most relevant and accurate information, regardless of any additional content provided.

\subsubsection{Reasoning Similarity Algorithm}
\paragraph{Step 1: Subject Matching.}
For each evidence triple, we first assess subject matching with response triples. Response triples with non-matching subjects are excluded from calculation.

\paragraph{Step 2: Predicate and Object Similarity.}
After subject matching, we evaluate predicate and object similarities according to the following cases:

\paragraph{Case 1: Similar Predicates}
\begin{itemize}
   \item When predicates are similar, use the predicate similarity score as the triple's similarity score
   \item This applies regardless of object similarity
\end{itemize}

\paragraph{Case 2: Dissimilar Predicates}
\begin{itemize}
   \item When predicates are not similar:
   \begin{itemize}
       \item If objects are similar: Use predicate similarity as the triple's score
       \item If objects are not similar: Exclude triple from analysis
   \end{itemize}
\end{itemize}

\paragraph{Step 3: Final Score Calculation. }
\begin{itemize}
   \item Calculate similarity scores for each evidence triple using the above criteria
   \item The final reasoning similarity score is the average of all valid triple similarity scores
\end{itemize}

\section{The Human Benchmark}\label{sec:human_benchmark}
\subsection{Question Selection}
To establish a human performance benchmark for comparison with LLM results, we selected a subset of 2,000 questions from our main dataset. The selection process involved careful manual verification to ensure that each question was unambiguous and that incorrect answers would stem from knowledge gaps or reasoning errors rather than linguistic confusion. 


\subsection{Experimental Protocol}
The data was collected through Prolific's online platform in January, 2025. Prior to the main task, participants received detailed instructions about the incentive mechanism, followed by a three-question comprehension test to verify their understanding of the procedure. Only participants who successfully passed this screening proceeded to the main experimental phase. \\
The core experimental task comprised ten independent rounds. In each round, participants responded to a binary-choice questions (requiring "Yes" or "No" responses) randomly selected from our validated question pool. Following each response, participants reported their confidence using a continuous scale ranging from 0 to 100, representing their subjective probability assessment of answer correctness.

\subsection{Incentive Scheme} \label{incentive_scheme}
The compensation structure was as follows:
Participants who failed the comprehension check received only the \$0.50 show-up fee and were excluded from the main task; If they complete the experiment, they would get \$ 2 as the base payment; Beyond this, they would get up to \$ 3  as the performance bonus based on the incentive mechanism following \citet{danz2022belief}. 

After each question, participants reported their confidence level, which determined their potential reward through the following mechanism:

For a randomly selected question, let $X \sim U[0,1]$ be a random draw and $c$ be the participant's reported confidence: if $X < c$: Participant receives \$5 if their answer is correct; otherwise participant receives \$5 with probability $X$. This mechanism ensures that reporting one's true confidence level is the optimal strategy so as to elicit truthful confidence reports.

\subsection{Summary Statistics}
For our primary analysis, we define two samples: one is the full sample (N=588), where all participants who completed the experiment; the other one is the baseline sample, where participants who complete the experiment and reported confidence $\geq 50\%$ for more than five out of ten questions (N=545).

Table \ref{tab:002_benchmark_comparison_summary} presents human performance by two measures across 10 rounds. "Diff." represents the difference between baseline sample and full sample and standard deviations are reported in parentheses. It shows no significant differences in performance patterns between the full and baseline samples, suggesting that our baseline criterion effectively identifies participants who engaged consistently with the confidence reporting task without introducing selection bias. In the main text, we use baseline samples to work as the benchmark of LLMs performance. 
\begin{table}[!h]
\centering
\caption{Human Benchmark Performance Across Rounds}
\centering
\begin{tabular}[t]{lcccccc}
\toprule
\multicolumn{1}{c}{ } & \multicolumn{3}{c}{Accuracy} & \multicolumn{3}{c}{Confidence} \\
\cmidrule(l{3pt}r{3pt}){2-4} \cmidrule(l{3pt}r{3pt}){5-7}
Round & Baseline & All & Diff. & Baseline & All & Diff.\\
\midrule
1 & 0.63 & 0.63 & -0.000 & 0.68 & 0.66 & 0.012\\
 &  &  &  & (0.212) & (0.220) & \\
\addlinespace
2 & 0.66 & 0.66 & 0.006 & 0.67 & 0.65 & 0.018\\
 &  &  &  & (0.225) & (0.235) & \\
\addlinespace
3 & 0.66 & 0.66 & -0.002 & 0.69 & 0.68 & 0.014\\
\addlinespace
 &  &  &  & (0.214) & (0.224) & \\
\addlinespace
4 & 0.68 & 0.67 & 0.007 & 0.69 & 0.67 & 0.015\\
 &  &  &  & (0.205) & (0.214) & \\
\addlinespace
5 & 0.65 & 0.65 & -0.003 & 0.69 & 0.67 & 0.015\\
 &  &  &  & (0.214) & (0.222) & \\
\addlinespace
\addlinespace
6 & 0.66 & 0.66 & 0.008 & 0.68 & 0.67 & 0.016\\
 &  &  &  & (0.220) & (0.229) \vphantom{1} & \\
\addlinespace
7 & 0.67 & 0.67 & 0.003 & 0.69 & 0.68 & 0.015\\
 &  &  &  & (0.212) & (0.222) & \\
\addlinespace
8 & 0.64 & 0.64 & 0.003 & 0.67 & 0.66 & 0.014\\
\addlinespace
 &  &  &  & (0.210) & (0.218) & \\
\addlinespace
9 & 0.64 & 0.64 & 0.006 & 0.68 & 0.66 & 0.016\\
 &  &  &  & (0.214) & (0.223) & \\
\addlinespace
10 & 0.66 & 0.66 & 0.002 & 0.69 & 0.67 & 0.015\\
 &  &  &  & (0.220) & (0.229) & \\
\addlinespace
\bottomrule
\label{tab:002_benchmark_comparison_summary}
\end{tabular}
\end{table}

Accordingly, Table \ref{tab:summary_human_confidence} presents a comparative analysis of confidence metrics between LLMs and human participants when responding to the same set of questions. 

\begin{table}
\centering
\begin{threeparttable}
    \caption{Summary of Accuracy, Confidence, and Bias}
\label{tab:summary_human_confidence}
\begin{tabular}{lcccr}
\toprule
 & \begin{tabular}[c]{c}Fraction of\\correct answers\end{tabular} & \begin{tabular}[c]{c}Confidence\\in answer\end{tabular} & Bias & $N$ \\
\midrule
GPT 3.5 & 0.35 (0.476) & 0.95 (0.043) & 0.60 (0.473) & 2000 \\ \addlinespace
GPT 4o & 0.68 (0.465) & 0.94 (0.056) & 0.26 (0.451) & 2000 \\ \addlinespace
GPT o1 & 0.81 (0.395) & 0.96 (0.048) & 0.15 (0.384) & 2000 \\ \addlinespace
Llama 3.1 & 0.62 (0.486) & 0.85 (0.113) & 0.24 (0.461) & 2000 \\ \addlinespace
Llama 3.2 & 0.62 (0.485) & 0.95 (0.075) & 0.33 (0.486) & 2000 \\ \addlinespace
Human Baseline & 0.66 (0.474) & 0.70 (0.203) & 0.04 (0.473) & 5220 \\ \addlinespace
\bottomrule
\end{tabular}
\begin{tablenotes}
    \item \textit{Notes:} SDs are in parentheses. All biases are significantly different from zero ($p < 0.001$) 
\end{tablenotes}
\end{threeparttable}
\end{table}

\clearpage
\section{The LLM-Exposure Experiment}\label{sec:llm_exposure_experiment}
\subsection{Experimental Procotol}
To examine how exposure to LLM outputs affects human decision-making and confidence calibration, we conducted a randomized controlled experiment with variations of exposure to LLM responses and their associated confidence levels.

We recruited participants through Prolific for a between-subjects experiment with three treatment conditions. The experiment proceeded in two stages. In Stage 1, all participants answered 10 questions randomly selected from the same pool  as the human benchmark of 2,000 YES/NO reasoning questions and provided confidence assessments for each answer. In stage 2, participants were randomly assigned at the individual level to one of three experimental conditions:

\begin{enumerate}
\item \textbf{Control:} Participants were presented with their Stage 1 responses as defaults and given the opportunity to revise their answers and confidence levels without additional information.

\item \textbf{LLM answer:} Participants viewed the LLM's answer (using either GPT-4o, GPT-o1, or Llama 3.1) before providing their revised answers and confidence assessments in the following format:

\textit{[GPT-4o / GPT-o1 / Llama 3.1] answered [Yes/No] to this question.}

\item \textbf{LLM answer with confidence:} Participants viewed both the LLM's answer and its stated confidence level before providing their revised answers and confidence assessments:

\textit{[GPT-4o / GPT-o1 / Llama 3.1] answered [Yes/No] to this question.}

\textit{[GPT-4o / GPT-o1 / Llama 3.1] indicated that its answer is correct with probability [x\%].}
\end{enumerate}

In each experimental round, the specific LLM model presented to participants was randomly selected from among GPT-4o, GPT-o1, and Llama 3.1. These models were selected based on their higher accuracy compared to other available models.

We used the same incentive structure shown in section \ref{incentive_scheme}. To ensure comprehension of the experimental procedures, participants who completed an example question with confidence assessment and answered three comprehension check questions about the instructions and incentive mechanism can enter the main part of the experiment.

\subsection{Summary Statistics}

There are 1364 participants who completed the experiment. We established our baseline sample by excluding participants who reported confidence exceeding 50\% for more than five of the ten questions (N=1161). Table \ref{tab:002_comparison_summary} presents accuracy rates and confidence levels from Stage 1, showing mean values with standard deviations in parentheses across all rounds. As shown in the table, no significant differences in accuracy were observed between the baseline and full sample conditions. All subsequent analyses are based on the baseline data.

\begin{table}[htbp]
\centering
\caption{Human Performance Across Rounds}
\begin{tabular}[t]{lcccccc}
\toprule
\multicolumn{1}{c}{ } & \multicolumn{3}{c}{Accuracy} & \multicolumn{3}{c}{Confidence} \\
\cmidrule(l{3pt}r{3pt}){2-4} \cmidrule(l{3pt}r{3pt}){5-7}
Round & Baseline & All & Diff. & Baseline & All & Diff.\\
\midrule
1 & 0.62 & 0.61 & 0.008 & 0.66 & 0.62 & 0.035\\
 
 &  &  &  & (0.212) & (0.232) & \\
\addlinespace
 
2 & 0.64 & 0.62 & 0.016 & 0.66 & 0.63 & 0.038\\
 
 &  &  &  & (0.215) & (0.235) & \\
\addlinespace
 
3 & 0.62 & 0.62 & -0.002 & 0.68 & 0.64 & 0.035\\
 
 &  &  &  & (0.206) & (0.228) & \\
\addlinespace
 
4 & 0.63 & 0.62 & 0.015 & 0.67 & 0.64 & 0.039\\
 
 &  &  &  & (0.200) & (0.228) & \\
\addlinespace
 
5 & 0.65 & 0.64 & 0.008 & 0.68 & 0.64 & 0.041\\
 
 &  &  &  & (0.204) & (0.229) & \\
\addlinespace
 
6 & 0.64 & 0.62 & 0.017 & 0.68 & 0.64 & 0.043\\
 
 &  &  &  & (0.201) & (0.226) & \\
\addlinespace
 
7 & 0.63 & 0.63 & 0.003 & 0.68 & 0.64 & 0.039\\
 
 &  &  &  & (0.193) & (0.222) & \\
\addlinespace
 
8 & 0.63 & 0.63 & 0.007 & 0.69 & 0.65 & 0.037\\
 
 &  &  &  & (0.196) & (0.220) & \\
\addlinespace
 
9 & 0.64 & 0.64 & 0.009 & 0.69 & 0.65 & 0.038\\
 
 &  &  &  & (0.203) & (0.225) & \\
\addlinespace
 
10 & 0.64 & 0.63 & 0.008 & 0.69 & 0.65 & 0.041\\
 
 &  &  &  & (0.197) & (0.223) & \\
\addlinespace
\bottomrule
\label{tab:002_comparison_summary}
\end{tabular}
\end{table}
 
Table \ref{tab:002_human_exp_summary_stat} presents summary statistics of pre-intervention accuracy and confidence across experimental conditions. We conducted one-way ANOVA tests on the first round baseline data to examine potential pre-existing differences among the experimental groups. For pre-intervention correctness, we found no significant differences among the three groups ($F(2, 1158) = 0.036$, $p = 0.964$). Similarly, for pre-intervention confidence, no significant differences were observed ($F(2, 1158) = 0.609$, $p = 0.544$). These results confirm successful randomization, with all groups demonstrating comparable baseline performance before experimental interventions were introduced.
\begin{table}
\centering
\caption{Accuracy Rates and Confidence Levels by Group Assignment}
\centering
\begin{tabular}[t]{lccccc}
\toprule
\multicolumn{1}{c}{ } & \multicolumn{2}{c}{Accuracy Rate} & \multicolumn{2}{c}{Confidence Level} & \multicolumn{1}{c}{ } \\
\cmidrule(l{3pt}r{3pt}){2-3} \cmidrule(l{3pt}r{3pt}){4-5}
  & Pre-test & Post-test & Pre-test & Post-test & $N$\\
\midrule
Control & 0.648 & 0.652 & 0.681 & 0.692 & 3960\\
 
 &  &  & (0.203) & (0.200) & \\
\addlinespace
 
LLM answer & 0.630 & 0.690 & 0.667 & 0.777 & 3860\\
 
 &  &  & (0.200) & (0.185) & \\
\addlinespace
 
LLM answer+ confidence & 0.624 & 0.699 & 0.685 & 0.845 & 3790\\
 
 &  &  & (0.207) & (0.159) & \\
\addlinespace
\bottomrule
\label{tab:002_human_exp_summary_stat}
\end{tabular}
\end{table}

\begin{figure}[htbp]
    \centering
    \includegraphics[width=\textwidth]{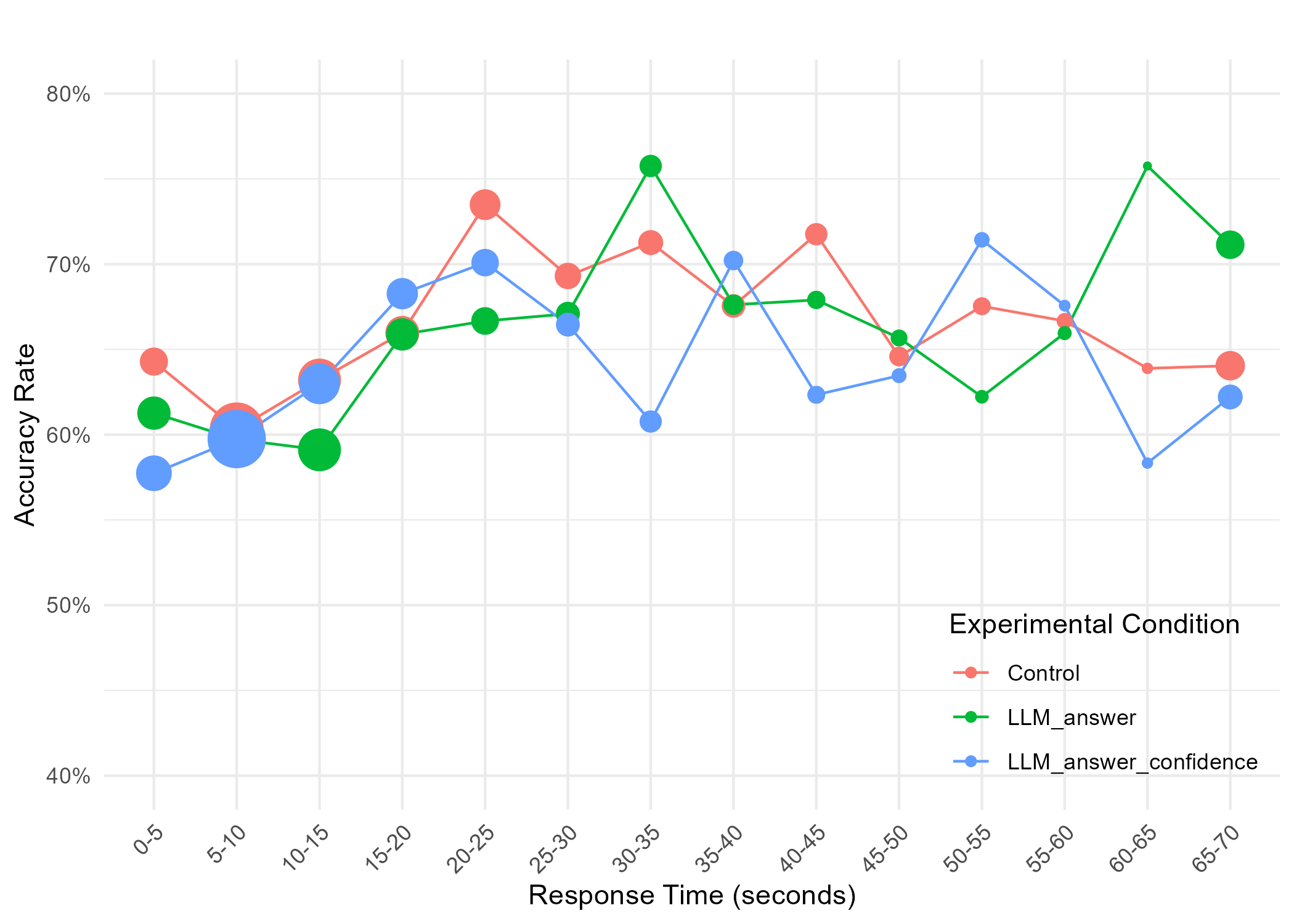}
    \caption{Average Accuracy Rate and Response Time across Experimental Conditions}
    \label{fig:response_time}
\end{figure}

Having established baseline equivalence across experimental conditions, we next examine participants' behavioral patterns during task completion. Figure \ref{fig:response_time} illustrates the relationship between participants' response times and accuracy rates across three experimental conditions. Response times were winsorized at the 95th percentile (70 seconds) and aggregated into 5-second intervals. The size of data points reflects the relative frequency of observations within each bin. All three conditions demonstrate comparable patterns, with accuracy rates generally fluctuating between 60\% and 75\% across the response time spectrum. When analyzing all 10 rounds collectively, the LLM answer with confidence group showed significantly faster pre-intervention response times compared to the other conditions (F(2, 11560) = 35.61, p < 0.001). However, our analysis of just the first round data revealed no significant differences in response times between groups (F(2, 1154) = 0.855, p = 0.425), again confirming successful randomization at baseline. This suggests that the observed response time differences emerged gradually over subsequent rounds rather than being present initially, indicating that repeated exposure to LLM answers and confidence information may progressively accelerate the answering process.

\begin{figure}[htbp]
    \centering
    \includegraphics[width=\textwidth]{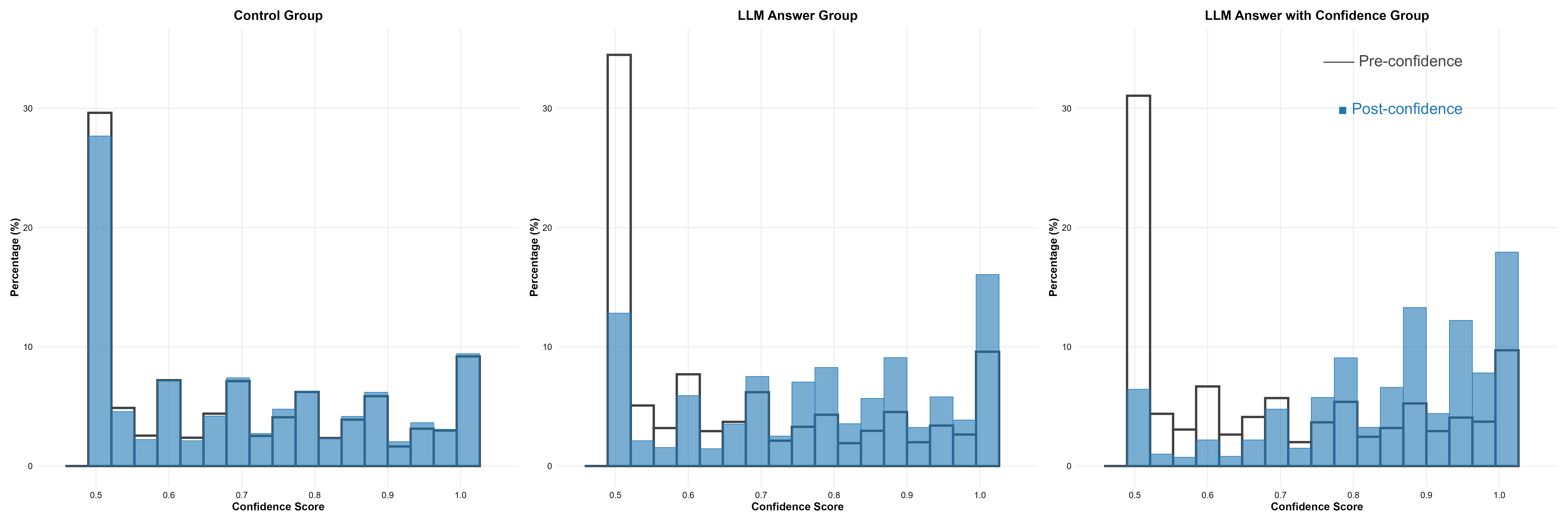}
    \caption{Confidence Distribution (Baseline and Intervention Stage)}
    \label{fig:conf_distribution_pre_post}
\end{figure}

While response time patterns reveal behavioral changes in how participants approached the tasks, our analysis of confidence calibration provides further insight into how LLM exposure influenced participants' cognitive assessments of their own performance. Figure \ref{fig:conf_distribution_pre_post} reveals significant changes in confidence distributions between baseline (outlined black bars) and intervention stages (blue filled bars) across the three experimental conditions. In both treatment conditions (LLM answer and LLM answer with confidence), we observe a pronounced rightward shift from lower to higher confidence ranges following intervention. This shift is particularly notable in the group with exposure to both LLM-generated answers and confidence, in which it appears to substantially increase participants' self-reported confidence levels. 

\subsection{Subsample Analysis of Treatment Effects}\label{subsec:subsample_analysis}

We conduct a subsample analysis to investigate heterogeneous treatment effects based on confidence levels. We categorize observations along two dimensions: LLM confidence and human baseline confidence. For LLM confidence, we define "high" as instances where the selected LLM's confidence exceeds 0.9, and "low" otherwise. For participants in the control group (who received no LLM assistance), we use the average confidence level across the three LLMs as the benchmark. For human baseline confidence, we classify observations as "high" when a participant's confidence at the baseline stage exceeds the sample median, and "low" otherwise.

This classification yields four distinct subgroups: (1) low LLM confidence, low human confidence; (2) low LLM confidence, high human confidence; (3) high LLM confidence, low human confidence; and (4) high LLM confidence, high human confidence.

Table \ref{tab:subsample_combined_summary} presents summary statistics for these four subgroups. Panel A shows the distribution of observations, which is not uniform across categories, with the largest proportion (33.9\%) in the high LLM confidence, high human confidence category. Panels B and C report human accuracy and bias at baseline, respectively, while Panels D and E present LLM accuracy and bias. Notably, LLM accuracy (Panel D) is substantially higher when LLM confidence is high (76.5-79.8\%) compared to when it is low (50.1-51.8\%), indicating that LLM confidence is a reliable predictor of accuracy.

\begin{table}
\centering
\caption{Combined Summary of LLM and Human Performance}
\centering
\label{tab:subsample_combined_summary}
\begin{tabular*}{\textwidth}{@{\extracolsep{\fill}}cccc@{}}
\toprule
\multicolumn{2}{c}{ } & \multicolumn{2}{c}{Human Confidence} \\
\cmidrule(l{3pt}r{3pt}){3-4}
  & LLM Confidence & Low & High\\
\midrule
\addlinespace[1em]
\multicolumn{4}{l}{\textbf{Panel A: Observation Percentages}}\\
\hspace{1em} & Low & 19.3\% & 16.2\%\\
\hspace{1em} & High & 29.4\% & 33.9\%\\
\addlinespace[1em]
\multicolumn{4}{l}{\textbf{Panel B: Human Accuracy}}\\
\hspace{1em} & Low & 0.543 (0.498) & 0.677 (0.468)\\
\hspace{1em} & High & 0.545 (0.498) & 0.745 (0.436)\\
\addlinespace[1em]
\multicolumn{4}{l}{\textbf{Panel C: Human Bias}}\\
\hspace{1em} & Low & 0.493 (0.097) & 0.37 (0.325)\\
\hspace{1em} & High & 0.49 (0.098) & 0.314 (0.325)\\
\addlinespace[1em]
\multicolumn{4}{l}{\textbf{Panel D: LLM Accuracy}}\\
\hspace{1em} & Low & 0.501 (0.457) & 0.518 (0.446)\\
\hspace{1em} & High & 0.765 (0.401) & 0.798 (0.376)\\
\addlinespace[1em]
\multicolumn{4}{l}{\textbf{Panel E: LLM Bias}}\\
\hspace{1em} & Low & 0.467 (0.324) & 0.445 (0.317)\\
\hspace{1em} & High & 0.247 (0.368) & 0.215 (0.346)\\
\bottomrule
\end{tabular*}
\end{table}


Table \ref{tab:subsample_correlation_matrix} presents the correlation matrix between LLM accuracy, LLM confidence, baseline human accuracy, and baseline human confidence. A key observation is that the correlation between LLM confidence and human baseline confidence is quite modest (0.082), suggesting that LLM and humans tend to be uncertain about different questions. This low correlation has important welfare implications: individuals who are uncertain about an answer typically receive LLM inputs of average quality, which generally exceed the accuracy of humans who are uncertain at baseline. If uncertainty were more strongly correlated between humans and LLMs, we would expect LLM assistance to yield smaller accuracy improvements for low-confidence humans while potentially increasing bias.

\begin{table}[htbp]
  \centering
    \caption{Correlation Matrix of LLM and Human Baseline Performance}
  \label{tab:subsample_correlation_matrix}
  \resizebox{\linewidth}{!}{
  \begin{tabular}{lcccc}
    \hline
     & LLM Accuracy & LLM Confidence & Baseline Accuracy & Baseline Confidence \\
    \hline
    LLM Accuracy &  & & & \\ \addlinespace
    LLM Confidence & 0.289$^{***}$ & & & \\ \addlinespace
    Baseline Accuracy & 0.182$^{***}$ & 0.034$^{***}$ &  & \\\addlinespace
    Baseline Confidence & 0.061$^{***}$ & 0.082$^{***}$ & 0.214$^{***}$ &\\\addlinespace
    \hline
  \end{tabular}}
  \begin{tablenotes}
    \small
    \item \textit{Notes:} $^{***}$ $p<0.001$, $^{**}$ $p<0.01$, $^{*}$ $p<0.05.$
  \end{tablenotes}
\end{table}

Table \ref{tab:009_subsample_treatment_effects} presents the treatment effects across the four subgroups. The reference category is the group with high LLM confidence and high human baseline confidence. With the main treatment effects maintaining the positive impacts on accuracy, we observe significant negative interaction effects on accuracy (-8.1 to -8.9 percentage points) and substantial positive effects on bias (+20.8 to +22.4 percentage points) when exposed to LLM input for the low LLM confidence, low human confidence subgroup.  This suggests that when both the human and LLM are uncertain, providing LLM input can be counterproductive.

\begin{table}[htbp]
   \caption{Subsample Treatment Effects across Experimental Conditions}
   \medskip
   \centering
   \resizebox{\linewidth}{!}{%
   \begin{tabular}{lcc}
      \toprule
                                                                    & $\Delta$Accuracy  & $\Delta$Bias\\   
      \addlinespace
      \midrule 
      LLM Answer                                                    & 0.051$^{***}$     & -0.020\\   
                                                                    & (0.011)           & (0.012)\\   
      \addlinespace
      LLM Answer+Confidence                                         & 0.056$^{***}$     & 0.005\\   
                                                                    & (0.012)           & (0.012)\\   
      \addlinespace
      Low LLM conf, Low human conf                                  & -0.010            & 0.025$^{**}$\\   
                                                                    & (0.007)           & (0.008)\\   
      \addlinespace
      High LLM conf, Low human conf                                 & 0.002             & 0.015$^{*}$\\   
                                                                    & (0.006)           & (0.007)\\   
      \addlinespace
      Low LLM conf, High human conf                                 & -0.002            & 0.001\\   
                                                                    & (0.005)           & (0.005)\\   
      \addlinespace
      LLM Answer $\times$ Low LLM conf, Low human conf              & -0.089$^{**}$     & 0.208$^{***}$\\   
                                                                    & (0.027)           & (0.029)\\   
      \addlinespace
      LLM Answer+Confidence $\times$ Low LLM conf, Low human conf   & -0.081$^{**}$     & 0.224$^{***}$\\   
                                                                    & (0.031)           & (0.032)\\   
      \addlinespace
      LLM Answer $\times$ High LLM conf, Low human conf             & 0.100$^{***}$     & 0.026\\   
                                                                    & (0.021)           & (0.023)\\   
      \addlinespace
      LLM Answer+Confidence $\times$ High LLM conf, Low human conf  & 0.156$^{***}$     & 0.080$^{***}$\\   
                                                                    & (0.022)           & (0.022)\\   
      \addlinespace
      LLM Answer $\times$ Low LLM conf, High human conf             & -0.100$^{***}$    & 0.091$^{***}$\\   
                                                                    & (0.021)           & (0.023)\\   
      \addlinespace
      LLM Answer+Confidence $\times$ Low LLM conf, High human conf  & -0.095$^{***}$    & 0.042\\   
                                                                    & (0.025)           & (0.026)\\   
      \addlinespace
      Constant                                                      & 0.005             & -0.0002\\   
                                                                    & (0.002)           & (0.003)\\   
      \addlinespace
      \midrule 
      Observations                                                  & 11,477            & 11,477\\  
      R$^2$                                                         & 0.03673           & 0.03143\\  
      \bottomrule
   \end{tabular}
   }
   
   \label{tab:009_subsample_treatment_effects}
                    \medskip 
        \begin{tablenotes}
      \small 
      \item \textit{Notes:$^{***}p < 0.001$; $^{**}p < 0.01$; $^{*}p < 0.05$. }
      Standard errors in parentheses are clustered at the player and question levels. The reference category is the group with high LLM confidence and high human baseline confidence.
   \end{tablenotes}
\end{table}

\subsection{Gender Gap in Overconfidence}

Our analysis reveals gender differences in confidence that align with well-documented patterns in prior literature (see 
\citealt{bordalo2019beliefshong} and \citealt{exley2022gender}). Table \ref{tab:008_gender_summary} presents a comparative analysis of accuracy and confidence measures by gender across all experimental conditions, including baseline values, post-exposure values, and the magnitude of changes between stages.

The data demonstrate a consistent gender gap in confidence levels but not in accuracy. Across all experimental groups, male participants report significantly higher baseline confidence than female participants. These confidence differences exist despite comparable accuracy levels between genders in most conditions. 

In response to LLM assistance, we observe notable gender differences in adaptability. In the LLM Answer group, women show larger confidence increases after exposure to LLM outputs compared to men (12.1 vs. 10.0 percentage points). In the LLM Answer with Confidence group, women demonstrate substantially larger accuracy improvements (9.1 percentage points) compared to men (6.0 percentage points), nearly eliminating the initial accuracy gap. The Control group shows minimal changes for both genders, confirming that observed effects in treatment groups are attributable to LLM assistance.

\begin{table}

\caption{Baseline Sample Summary Table by Gender and Groups}
\centering
\begin{tabular}[t]{lcccc}
\toprule
Measure & Female & Male & Difference & p.value\\
\midrule
\addlinespace[0.3em]
\multicolumn{5}{l}{\textbf{LLM Answer Group}}\\
\hspace{1em}Baseline Accuracy & 0.635 & 0.625 & -0.009 & 0.544\\
\addlinespace
\hspace{1em}Post-exposure Accuracy & 0.693 & 0.685 & -0.008 & 0.598\\
\addlinespace
\hspace{1em}Accuracy Change & 0.058 & 0.060 & 0.002 & 0.921\\
\addlinespace
\hspace{1em}Baseline Confidence & 0.642 & 0.688 & 0.046 & 0.000\\
\addlinespace
\hspace{1em}Post-exposure Confidence & 0.763 & 0.788 & 0.026 & 0.000\\
\addlinespace
\hspace{1em}Confidence Change & 0.121 & 0.100 & -0.020 & 0.000\\
\addlinespace
\addlinespace[0.3em]
\multicolumn{5}{l}{\textbf{LLM Answer with Confidence Group}}\\
\hspace{1em}Baseline Accuracy & 0.606 & 0.641 & 0.035 & 0.025\\
\addlinespace
\hspace{1em}Post-exposure Accuracy & 0.696 & 0.701 & 0.005 & 0.742\\
\addlinespace
\hspace{1em}Accuracy Change & 0.091 & 0.060 & -0.031 & 0.074\\
\addlinespace
\hspace{1em}Baseline Confidence & 0.669 & 0.700 & 0.031 & 0.000\\
\addlinespace
\hspace{1em}Post-exposure Confidence & 0.829 & 0.859 & 0.030 & 0.000\\
\addlinespace
\hspace{1em}Confidence Change & 0.160 & 0.159 & -0.001 & 0.892\\
\addlinespace
\addlinespace[0.3em]
\multicolumn{5}{l}{\textbf{Control Group}}\\
\hspace{1em}Baseline Accuracy & 0.654 & 0.644 & -0.010 & 0.491\\
\addlinespace
\hspace{1em}Post-exposure Accuracy & 0.658 & 0.646 & -0.012 & 0.411\\
\addlinespace
\hspace{1em}Accuracy Change & 0.004 & 0.002 & -0.002 & 0.682\\
\addlinespace
\hspace{1em}Baseline Confidence & 0.670 & 0.689 & 0.018 & 0.005\\
\addlinespace
\hspace{1em}Post-exposure Confidence & 0.683 & 0.700 & 0.017 & 0.007\\
\addlinespace
\hspace{1em}Confidence Change & 0.013 & 0.011 & -0.001 & 0.589\\
\addlinespace
\bottomrule
\label{tab:008_gender_summary}
\end{tabular}
\end{table}

These results suggest that LLM assistance, particularly when accompanied by explicit confidence information, may help reduce gender-based differences in decision-making performance. The greater responsiveness of female participants to LLM inputs represents a promising avenue for using AI systems to potentially mitigate pre-existing confidence gaps.

\section{The Results from the Exhibits in the Main Text}
\subsection{Table 1: Accuracy, confidence, and bias across models}
\subsubsection{Description of Table 1}
Table \ref{tab:desc_stats} presents descriptive statistics for key variables across five Large Language Models (LLMs), based on their performance on 10,000 reasoning problems.There are a small proportion of missing values resulting from either premature termination of LLM outputs or instances where models expressed confidence in textual rather than numerical format. For each measure, we report both the mean value and its standard deviation (in parentheses).\\
Section (a) reports the fraction of correct answers, representing each model's accuracy rate across the full sample. The results show substantial variation in performance across models,  indicating meaningful differences in reasoning capabilities across models. \\
Section (b) presents self-reported confidence in the overall answer correctness ($\tilde{p}$).  Notably, all models exhibit consistently high confidence, with most values exceeding 0.90, despite considerable variations in their actual performance. \\
Section (c) quantifies bias using difference between self-reported confidence ($\tilde{p}$) and actual accuracy. There is substantial positive bias across all models, indicating systematic overconfidence. Table \ref{tab:overconfidence_test} test whether bias are significantly different from zero.  Panel A shows the estimates using self-reported confidence, while Panel B presents the results using our derived confidence measure  ($\widetilde{p}_{F\cdot R}$) that combines the model's confidence in facts and reasoning through multiplication ($p_F \cdot p_R$). We constructed this measure to address potential conjunction fallacies in confidence assessment, providing a more conservative estimate of overall confidence. The derived measure consistently yields lower values than the direct self-reported confidence in answers, suggesting that decomposing confidence assessment into constituent components may help mitigate overconfidence. The coefficient estimates (labeled "Overconfidence") represent the average bias for each model, with robust standard errors reported in parentheses.

The results provide strong statistical evidence of systematic overconfidence across all models regardless of whether we measure it using self-reported or derived confidence. All coefficients in these two subtables are positive and highly statistically significant ($p < 0.001$), indicating that every model systematically overestimates its probability of being correct.

\begin{table}[ht]
\centering
\caption{Overconfidence Tests in Self-Reported and Derived Confidence}
\label{tab:overconfidence_test}
\begin{subtable}{\textwidth}
\centering
\begin{tabular}{lccccc}
\addlinespace
\multicolumn{6}{c}{Panel A: Self-Reported Confidence} \\ 
\toprule
& gpt 3.5 & gpt 4o & gpt o1 & Llama 3.1 & Llama 3.2  \\
\midrule
Overconfidence & 0.59 & 0.30 & 0.22 & 0.23 & 0.33 \\
& (0.005) & (0.005) & (0.004) & (0.005) & (0.005) \\
\addlinespace
$N$& 9,795 & 9,973 & 9,990 & 9,750 & 9,910 \\
\bottomrule
\end{tabular}
\end{subtable}
\hfill
\vspace{1em}
\begin{subtable}{\textwidth}
\centering
\begin{tabular}{lccccc}
\multicolumn{6}{c}{Panel B: Derived Confidence} \\ 
\toprule
& gpt 3.5 & gpt 4o & gpt o1 & Llama 3.1 & Llama 3.2  \\
\midrule
Overconfidence & 0.52 & 0.25 & 0.17 & 0.16 & 0.24 \\
& (0.005) & (0.005) & (0.004) & (0.005) & (0.005) \\
\addlinespace
$N$ & 9,779 & 9,895 & 9,974 & 9,614 & 9,854 \\
\bottomrule
\end{tabular}
\end{subtable}

\begin{tablenotes}
\small 
\item \textit{Notes:} All overconfidence measures are significantly different from zero at p<0.001. Confidence levels are normalized by subtracting 1 from the original confidence scores.
\end{tablenotes}
\end{table}

\clearpage

\subsection{Figure 1: LLM Confidence and Accuracy}
\subsubsection{Description of Figure 1}
Figure \ref{fig:confidence_analysis} presents the visualization of the relationship between confidence and accuracy across LLM models and the human benchmark.

In all panels of Figure \ref{fig:confidence_analysis}, the horizontal axis represents the self-reported confidence levels and the vertical axis represents the mean accuracy rate conditional at each confidence level. The 45-degree line represents perfect calibration, where confidence exactly matches accuracy. Points below this line indicate overconfidence, while points above would indicate underconfidence. The size of each plotted point is proportional to the number of observations at that confidence level. 

Panel A focuses on the GPT family of models (GPT 3.5, GPT 4o, and GPT o1), while Panel B presents the corresponding analysis for Llama models (Llama 3.1 8B and Llama 3.2 3B), both with 10,000 observations. Further, Panel C provides a comparative analysis between advanced LLM models (GPT 4o, GPT o1, and Llama 3.1 8B) and human performance, based on the human benchmark subset of 2,000 questions. 

The fitted lines represent weighted regression coefficients, where observations are weighted by their frequency. The shaded areas around each fitted line represent 95\% confidence bands, indicating the precision of our estimates. All slope coefficients (coefficients correspond to Panel A and B are detailed in Table \ref{tab:reg_acc_conf_10000} and those for Panel C are detailed in Table \ref{tab:sub_reg_conf_acc}) are statistically significant ($p < 0.001$) across models and comparison groups, suggesting robust relationships between confidence and accuracy, albeit with systematic overconfidence. 

\subsubsection{Comparison of Confidence Measures}
 To examine whether the patterns observed with self-reported confidence persist when using a measure that potentially mitigates conjunction fallacy, we also conducted parallel analyses on LLMs using our derived confidence measure ($\widetilde{p}_{F\cdot R}$) in Table \ref{tab:acc_derived_conf_10000}.

\begin{table}[!htbp]
    \centering
    \begin{threeparttable}
       \caption{Derived Confidence gradients in LLM accuracy}
       \label{tab:acc_derived_conf_10000}
   \begin{tabular}{lccccc}
        \toprule \addlinespace
        & gpt 3.5 & gpt 4o & gpt o1 & llama3.1 & llama3.2 \\
        \addlinespace \midrule \addlinespace
        Confidence & 0.33 & 1.52& 1.57& 0.85& 0.56\\
        gradient & (0.06) & (0.06) & (0.07) & (0.04) & (0.04) \\ \addlinespace
        Predicted accuracy & 0.39& 0.81& 0.89& 0.80& 0.70\\
        if $\widetilde{p}_{F\cdot R}=1$ & (0.01) & (0.01) & (0.01) & (0.01) & (0.01) \\
        \addlinespace \midrule \addlinespace
        $N$ & 9,779 & 9,895 & 9,974 & 9,614 & 9,854 \\
        R$^2$ & 0.003 & 0.07 & 0.06 & 0.05 & 0.02 \\
        \addlinespace \bottomrule
    \end{tabular}
\medskip 
\begin{tablenotes}
\small 
\item
\textit{Notes:}  Dependent variable is correct answer (= 1). Confidence levels are normalized by subtracting 1 from the original confidence scores, so the constant term represents accuracy at 100\% confidence ($\tilde{p} = 1$). All coefficients are significant at p<0.001.
\end{tablenotes}
\end{threeparttable}
\end{table}

\begin{figure}[!htbp]
    \centering
    \includegraphics[width=1.1\textwidth]{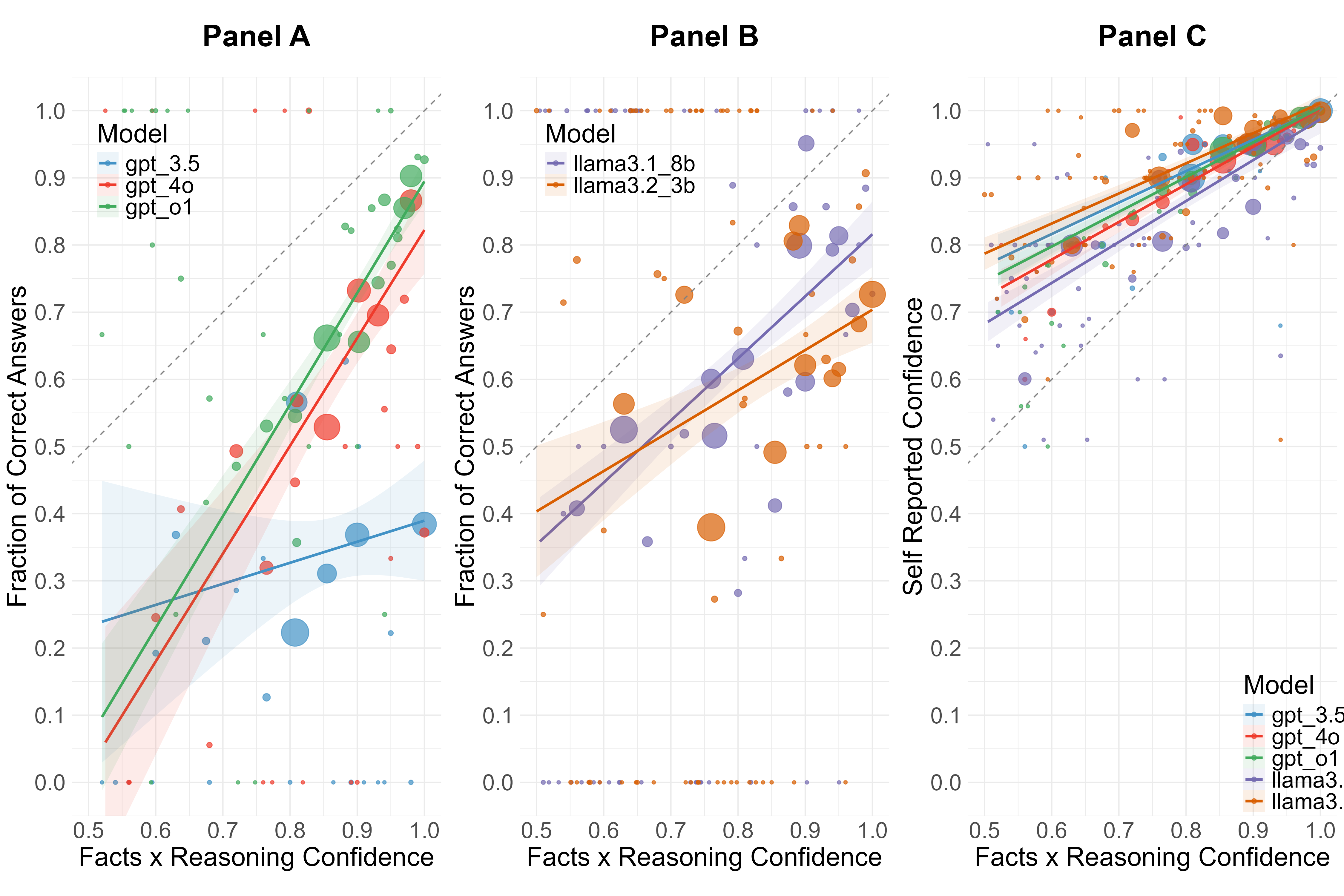}
    \caption{LLM Derived Confidence and Accuracy}
    \label{fig:acc_derive_conf}
\end{figure}

The analysis reveals similar qualitative patterns across both GPT and Llama model families, as illustrated in Panels A and B of Figure \ref{fig:acc_derive_conf}. However, Panel C reveals an important nuance in the relationship between self-reported and derived confidence measures. While the two measures are highly correlated, self-reported confidence consistently exceeds the derived measure, providing evidence of conjunction bias in model self-assessment. Notably, the difference between self-reported and derived confidence narrows at higher confidence levels, suggesting that models' assessments become more internally consistent—and potentially more rational—when they are most confident in their answers.

\subsection{Table 2: The confidence gradient in accuracy}
\subsubsection{Description of Table 2}

Table \ref{tab:reg_acc_conf_10000} presents the regression analysis quantifying the relationship between accuracy and confidence across different models using the full sample of questions. For each source, we estimate:\\
$ \mathbf{1}_{\{correct,i\}} = \beta_0 + \beta_1(Confidence_i - 1) + \epsilon_i $

where $\mathbf{1}_{\{correct,i\}}$ is an indicator equal to 1 if answer $i$ is correct, and $Confidence_i$ represents either self-reported confidence ($\widetilde{p}$) for all sources or derived confidence ($\widetilde{p}_{F\cdot R}$) for models. The confidence measures are normalized by subtracting 1, so that $\beta_0$ directly represents predicted accuracy at full confidence.

The confidence gradient ($\beta_1$) quantifies how accuracy changes with confidence levels, corresponding to the values of the fitting lines of Panel A and B in Figure \ref{fig:confidence_analysis}. The predicted accuracy at full confidence ($\beta_0$) reveals systematic overconfidence. 
The regression estimates are weighted by the frequency of
observations at each confidence level and heteroskedasticity-consistent standard errors (HC0) are reported in parentheses. The results reveal substantial heterogeneity across models. More advanced models (such as GPT 4o and GPT o1) exhibit steeper confidence gradients, indicating a stronger relationship between their confidence assessments and actual performance. However, even these models show significant overconfidence at full confidence levels, as evidenced by their intercept terms being consistently below 1.

\subsubsection{Confidence Gradients on Human Benchmark Subset}
We extend our analysis by estimating confidence gradients for the five LLM models and human benchmark using the same set
 of 2000 questions. These estimates correspond to the fitted lines displayed in Panels C of Figures \ref{fig:confidence_analysis}.

Confidence levels are normalized by subtracting 1 from the original confidence scores, so the constant term represents accuracy at 100\% confidence ($\tilde{p} = 1$). These regression estimates are weighted by the frequency of observations at each confidence level. For human participants, standard errors are clustered at both individual and question levels. 

\begin{table}[h!]
\centering
\begin{threeparttable}
 \caption{Confidence Gradient in Accuracy (Human Baseline Subsample)} 
\label{tab:sub_reg_conf_acc}
\begin{tabular}{lcccccc}  
\toprule \addlinespace
& human & gpt 3.5 & gpt 4o & gpt o1 & llama3.1 & llama3.2 \\
\addlinespace \midrule \addlinespace
Confidence & 0.51 & 1.39 & 2.50 & 2.26 & 1.42 & 0.32 \\
gradient & (0.03) & (0.28) & (0.19) & (0.31) & (0.10) & (0.15) \\ \addlinespace
Predicted accuracy & 0.81 & 0.41 & 0.83 & 0.91 & 0.82 & 0.64 \\
if $\tilde{p} = 1$ & (0.01) & (0.02) & (0.01) & (0.01) & (0.02) & (0.01) \\
\addlinespace \midrule \addlinespace
$N$ & 5,220 & 1,933 & 1,994 & 1,997 & 1,958 & 1,984 \\
$R^2$ & 0.05 & 0.02 & 0.09 & 0.08 & 0.11 & 0.002 \\
\addlinespace \bottomrule
\end{tabular}
\medskip
\begin{tablenotes}
\small
\item \textit{Notes:} Dependent variable is correct answer (= 1). The table presents the gradient of accuracy with respect to self-reported confidence in answer correctness across models using the human baseline subsample. Confidence levels are normalized by subtracting 1 from the original confidence scores.  All coefficients are significant at p<0.001, except for Llama 3.2, which is significant at p<0.05.
\end{tablenotes}
\end{threeparttable}
\end{table}

\subsection{Figure 2: Treatment Effects on Changes in Accuracy and Bias }
\subsubsection{Description of Figure 2}
The key variables of interest in the LLM-exposure experiment are the differences in accuracy and bias (measured by the gap between self-reported confidence and actual correctness) between pre-treatment and post-treatment measurements.Table \ref{tab:004_1_treatment_effects} presents regression results across treatment conditions, where $\Delta$accuracy and $\Delta$bias represent the differences between baseline and intervention stages for each respective measure.

\begin{table}[htbp]
   \caption{Treatment Effects across Experimental Conditions}
   \label{tab:004_1_treatment_effects}
    \medskip 
   \centering
   \begin{tabular}{lcc}
      \toprule
                              & $\Delta$Accuracy  & $\Delta$Bias\\   
      \addlinespace
      \midrule 
      LLM Answer              & 0.056$^{***}$     & 0.042$^{***}$\\   
                              & (0.010)           & (0.011)\\   
      \addlinespace
      LLM Answer + Confidence  & 0.071$^{***}$     & 0.076$^{***}$\\   
                              & (0.010)           & (0.011)\\   
      \addlinespace
      Constant                & 0.003             & 0.009$^{***}$\\   
                              & (0.002)           & (0.003)\\   
      \addlinespace
      \midrule 
      Observations            & 11,610            & 11,610\\  
      R$^2$                   & 0.00530           & 0.00522\\  
      \bottomrule
   \end{tabular}

    \medskip 
    \begin{tablenotes}
      \small 
      \item \textit{Notes:} $^{***}p < 0.001$; $^{**}p < 0.01$; $^{*}p < 0.05$. 
      Standard errors in parentheses are clustered at the player and question levels.
   \end{tablenotes}
\end{table}

The regression results reveal substantial treatment effects across outcome variables. Both experimental conditions significantly improved accuracy rates, with LLM Answer+Confidence Group showing a more pronounced effect (7.1 percentage points) compared to the LLM Answer Group (5.6 percentage points) but there is no statistically significant difference in accuracy improvement between these two treatment groups ($\chi^2(1) = 1.55$, $p = 0.213$). The full treatment effects for each experimental condition are illustrated in the Panel A of Figure  \ref{fig:treatment_effects} in the main text. 

Notably, both treatments significantly increased participants' bias with the LLM Answer+Confidence Group exhibiting a stronger effect (7.6 percentage points) compared to the LLM Answer Group (4.2 percentage points) and this difference in bias induction between treatment groups was statistically significant ($\chi^2(1) = 6.46$, $p = 0.011$). The full treatment effects on bias are displayed in the Panel B of Figure \ref{fig:treatment_effects} in the main text.

\subsection{Table 3: Subsample Analysis of LLM Exposure Experiment}
In table \ref{tab:heter_treatment_effect}, we examine heterogeneity across two dimensions: (1) how the presence of LLM confidence information influences participants' judgment and decision-making, (2) how participants' initial confidence levels moderate the effects of LLM assistance. The  full regression results see Table \ref{tab:subsample_full_results} below.  This analysis provides insights into the contexts and populations for whom LLM assistance offers the greatest benefit.

\begin{table}[htbp]
   \caption{\textbf{Subsample Analysis of LLM Exposure Experiment - Full Results}}
   \medskip
   \centering
   \resizebox{\linewidth}{!}{%
   \begin{threeparttable}
   \begin{tabular}{lcccc}
      \toprule
       & \multicolumn{2}{c}{Human Baseline Confidence} & \multicolumn{2}{c}{LLM Confidence} \\ 
       \cmidrule(lr){2-3} \cmidrule(lr){4-5}
      \addlinespace
      Dependent variable: & $\Delta$ Accuracy  & $\Delta$ Bias  & $\Delta$ Accuracy  & $\Delta$ Bias \\    
      \midrule \addlinespace
      LLM Answer                                         & 0.086              & 0.070          & -0.033             & 0.133\\   
                                                         & (0.015)            & (0.017)        & (0.018)            & (0.020)\\   
      \addlinespace
      LLM Answer + Conf.                                   & 0.119              & 0.141          & -0.028             & 0.160\\   
                                                         & (0.017)            & (0.018)        & (0.020)            & (0.021)\\   
      \addlinespace
      High Confidence (Human)                            & 0.001              & -0.018         &                    &   \\   
                                                         & (0.005)            & (0.005)        &                    &   \\   
      \addlinespace
      LLM Answer $\times$ High Confidence (Human)        & -0.064             & -0.063         &                    &   \\   
                                                         & (0.018)            & (0.019)        &                    &   \\   
      \addlinespace
      LLM Answer + Conf. $\times$ High Confidence (Human)  & -0.091             & -0.124         &                    &   \\   
                                                         & (0.019)            & (0.019)        &                    &   \\   
      \addlinespace
      High Confidence (LLM)                              &                    &                & 0.008              & -0.008\\   
                                                         &                    &                & (0.005)            & (0.006)\\   
      \addlinespace
      LLM Answer $\times$ High Confidence (LLM)          &                    &                & 0.136              & -0.138\\   
                                                         &                    &                & (0.020)            & (0.021)\\   
      \addlinespace
      LLM Answer + Conf. $\times$ High Confidence (LLM)    &                    &                & 0.153              & -0.119\\   
                                                         &                    &                & (0.024)            & (0.024)\\   
      \addlinespace
      Constant                                           & 0.003              & 0.018          & -0.002             & 0.014\\   
                                                         & (0.004)            & (0.005)        & (0.004)            & (0.005)\\   
      \addlinespace
      \midrule 
      Observations                                       & 11,610             & 11,610         & 10,957             & 10,957\\  
      R$^2$                                              & 0.01082            & 0.01700        & 0.02477            & 0.01915\\  
      \bottomrule
   \end{tabular}
   \begin{tablenotes}
      \small 
      \item \textit{Notes:}  The table presents the full regression results including all control variables and interaction terms. The main text Table \ref{tab:heter_treatment_effect} presents a condensed version focusing on key treatment interactions.
      \item $^{***}p < 0.001$; $^{**}p < 0.01$; $^{*}p < 0.05$.
   \end{tablenotes}
   \end{threeparttable}
   }
   \label{tab:subsample_full_results}
\end{table}

\subsubsection{Conditional on LLM Confidence}
To investigate how LLM confidence levels moderate treatment effects, we constructed LLM confidence measures that varied by experimental condition. For the control group, we used the average confidence level from GPT-4o, GPT-o1, and Llama 3.1 on each question. For treatment groups (both LLM answer and LLM answer with confidence), we used the confidence level expressed by the specific model that participants encountered for each question. 

\begin{figure}
    \centering
    
    \begin{subfigure}[b]{\textwidth}
        \centering
        \includegraphics[width=0.8\textwidth]{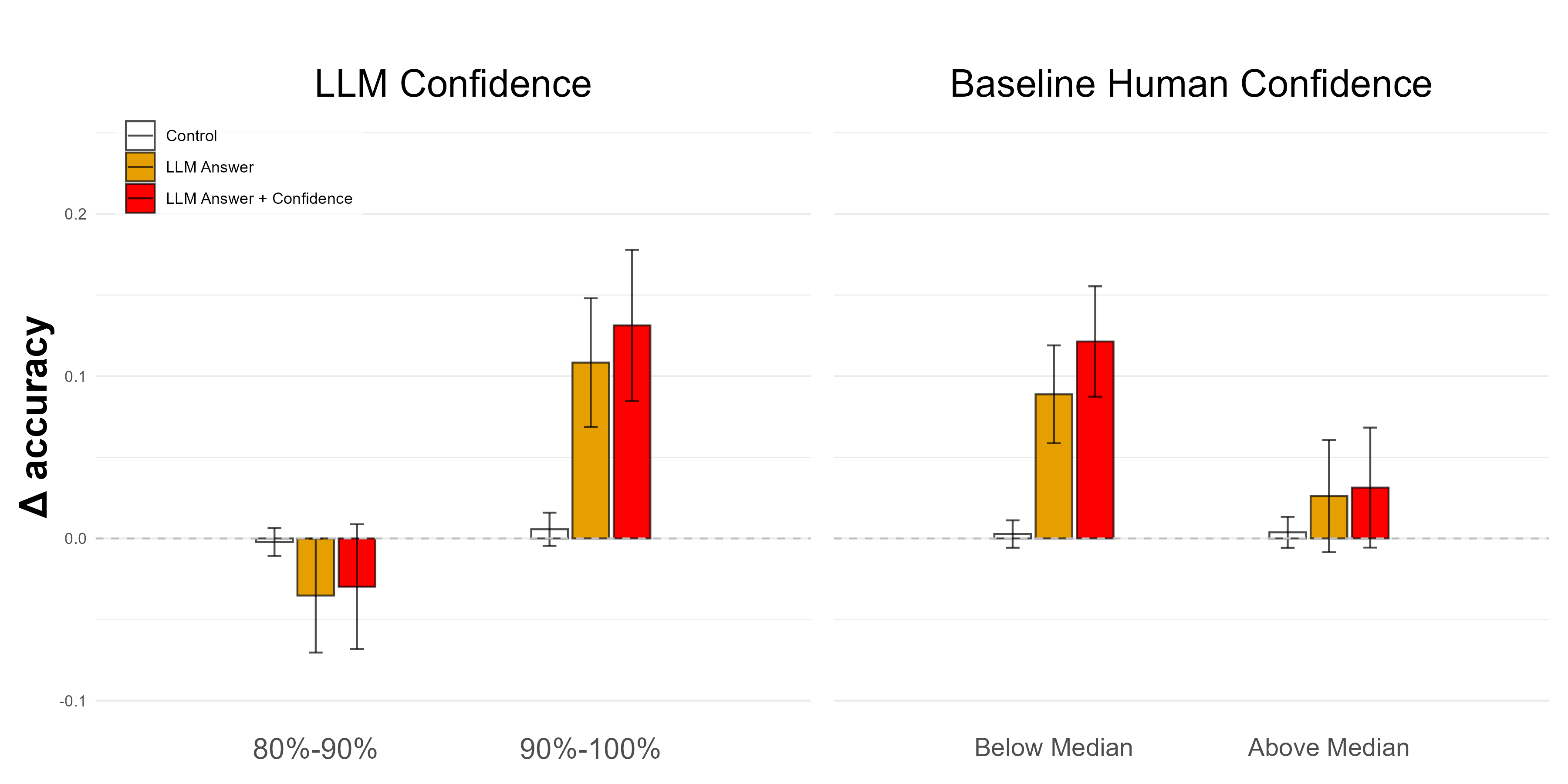}
        \caption*{Panel A: Heterogeneous Treatment Effects on Change in Accuracy}
    \end{subfigure}
    
    \vspace{1cm} 
    
    \begin{subfigure}[b]{\textwidth}
        \centering
        \includegraphics[width=0.8\textwidth]{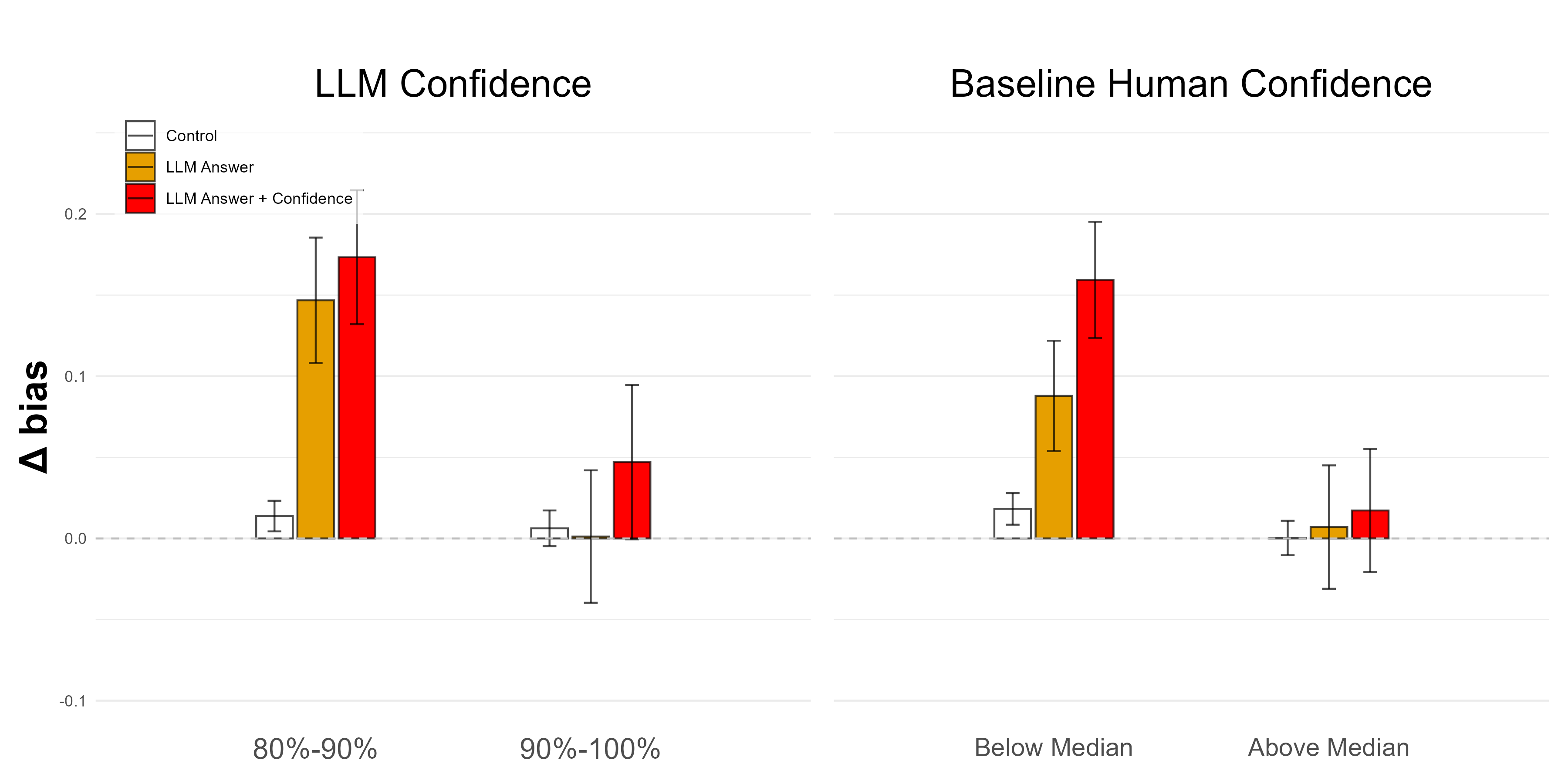}
        \caption*{Panel B: Heterogeneous Treatment Effects on Change in Bias}
        \label{fig:treatment_effect_bias}
    \end{subfigure}
    \caption{Illustration of Subsample Analysis}
    \label{fig:heter_treatment_effects}
\end{figure}

To examine whether high LLM confidence levels distinctly influence treatment effects, columns 1 and 2 of Table \ref{tab:heter_treatment_effect} present regression results using the subsample with LLM confidence above 0.8  (on a 0-1 scale). We define high confidence as a binary variable indicating LLM confidence scores above 0.9. The combined treatment effects with 95\% confidence intervals for each experimental condition are illustrated in the first part of Panel A and Panel B of Figure \ref{fig:heter_treatment_effects} separately. 

The findings reveal striking interaction effects. While the main treatment effects show minimal impact on accuracy changes, their interactions with high confidence demonstrate substantial effects. Specifically, when LLMs express high confidence (confidence ranging from 0.9 to 1), the LLM Answer treatment increases accuracy by 13.6 percentage points, while the LLM Answer with Confidence treatment produces an even larger improvement of 15.3 percentage points. Similarly, both treatments significantly reduce bias when coupled with high LLM confidence. The interaction between high confidence and the LLM Answer treatment reduces bias by 13.8 percentage points, while the interaction with the LLM Answer with Confidence treatment reduces bias by 11.9 percentage points. Such results show that the benefits of LLM assistance are concentrated in scenarios where the models express high confidence in their answers, suggesting that confidence signals contain valuable information that helps participants discriminate between reliable and unreliable AI outputs.

\subsubsection{Conditional on Human Baseline Confidence}

Understanding how initial user confidence mediates the effects of LLM assistance is also crucial for optimizing human-AI collaboration. Columns 3 and 4 of Table \ref{tab:heter_treatment_effect} present our heterogeneity analysis examining how participants' baseline confidence levels (i.e., their initial confidence at pre-treatment stage) moderate the effects of LLM answers on accuracy and bias, where high confidence is defined as above median baseline confidence. Accordingly, combined treatment effects are illustrated in the second part of Panel A and Panel B in Figure \ref{fig:heter_treatment_effects} separately. 

The results reveal substantial heterogeneity in treatment effects. Specifically, high-confidence participants show reductions of 6.4 percentage points in the accuracy benefit from LLM Answer and 9.1 percentage points in the benefit from LLM Answer+Confidence. The pattern is similar for bias reduction, with high-confidence participants showing reductions of 6.3 and 12.4 percentage points respectively in the treatment benefits.

These results suggest that LLM assistance provides the greatest value to users who lack confidence in their initial judgments, potentially because they are more receptive to incorporating AI suggestions. Conversely, highly confident users appear more resistant to revising their judgments based on LLM inputs, resulting in significantly smaller improvements from AI assistance.

\subsubsection{Subsample Analysis: Continuous Version}

To provide a more comprehensive picture of heterogeneous treatment effects and to serve as a robustness check for Table \ref{tab:heter_treatment_effect}, Table \ref{tab:004_3_continuous_heter_treatments} presents regression results examining how continuous measures of LLM confidence and human baseline confidence relate to changes in participant accuracy and bias across experimental conditions. 

\begin{table}[htbp]
   \caption{ Continuous Analysis of Treatment Effects Moderated by LLM and Human Confidence}
   \medskip
   \resizebox{\linewidth}{!}{%
   \centering
   \begin{tabular}{lcccc}
      \toprule
       & \multicolumn{2}{c}{Continuous LLM Confidence} & \multicolumn{2}{c}{Baseline Human Confidence} \\ \cmidrule(lr){2-3} \cmidrule(lr){4-5}
      \addlinespace
                                                       & $\Delta$ Accuracy  & $\Delta$ Bias  & $\Delta$ Accuracy  & $\Delta$ Bias \\    
      \midrule 
      LLM Answer                                       & -0.595$^{***}$     & 0.688$^{***}$  & 0.086$^{***}$      & 0.070$^{***}$\\   
                                                       & (0.101)            & (0.108)        & (0.014)            & (0.017)\\   
      \addlinespace
      LLM Answer+Confidence                            & -0.555$^{***}$     & 0.310$^{**}$   & 0.114$^{***}$      & 0.139$^{***}$\\   
                                                       & (0.098)            & (0.098)        & (0.016)            & (0.017)\\   
      \addlinespace
      LLM Confidence                                   & 0.053              & -0.053         &                    &   \\   
                                                       & (0.036)            & (0.041)        &                    &   \\   
      \addlinespace
      LLM Answer            & 0.707$^{***}$      & -0.702$^{***}$ &                    &   \\   
               $\times$ LLM Confidence                                            & (0.109)            & (0.115)        &                    &   \\   
      \addlinespace
      LLM Answer+Confidence   & 0.685$^{***}$      & -0.256$^{*}$   &                    &   \\   
           $\times$ LLM Confidence                                              & (0.106)            & (0.106)        &                    &   \\   
      \addlinespace
      Baseline Confidence                                 &                    &                & -0.010             & -0.050$^{***}$\\   
                                                       &                    &                & (0.012)            & (0.015)\\   
      \addlinespace
      LLM Answer             &                    &                & -0.177$^{***}$     & -0.174$^{***}$\\   
               $\times$ Baseline Confidence                                         &                    &                & (0.040)            & (0.045)\\   
      \addlinespace
      LLM Answer+Confidence     &   &                & -0.233$^{***}$     & -0.338$^{***}$\\   
        $\times$ Baseline Confidence                               &                    &                & (0.044)            & (0.047)\\   
      \addlinespace
      Constant                                         & -0.046             & 0.057          & 0.005              & 0.018$^{***}$\\   
                                                       & (0.033)            & (0.038)        & (0.004)            & (0.005)\\   
      \addlinespace
      \midrule 
      Observations                                     & 11,477             & 11,477         & 11,610             & 11,610\\  
      R$^2$                                            & 0.02181            & 0.01409        & 0.01256            & 0.02005\\    
      \bottomrule
   \end{tabular}
   }
   
   \medskip \label{tab:004_3_continuous_heter_treatments}
      \textit{Notes:} Models 1-2 include LLM confidence (mean-centered at 0.8), and Models 3-4 use respondents' prior confidence as a moderator. \\
     $^{***}p < 0.001$; $^{**}p < 0.01$; $^{*}p < 0.05$.\\
     Standard errors in parentheses are clustered at the participant and question levels.
\end{table}

For LLM confidence, our findings reveal significant patterns. While main treatment effects on accuracy are slightly negative, these effects are substantially moderated by LLM confidence levels. The interaction between LLM confidence and both treatment conditions shows strong positive effects on accuracy changes and corresponding negative effects on bias changes. This indicates that when LLMs express higher confidence, their answers produce greater improvements in participant accuracy, though bias reduction may be less pronounced when LLM confidence is explicitly shared.

When examining human baseline confidence, participants shown LLM answers experience accuracy improvements of 8.6 percentage points compared to controls, increasing to 11.9 percentage points when LLM confidence is also provided. Similar patterns emerge for bias reduction (7.0 and 13.9 percentage points respectively). However, the significant negative interaction coefficients demonstrate that participants with higher baseline confidence benefit substantially less from LLM assistance, with this effect being even more pronounced for bias reduction.

These continuous variable analyses confirm the patterns observed in our subsample analysis, providing robust evidence that LLM assistance is most beneficial when models express high confidence and when users initially express low confidence in their judgments.

\subsection{Descriptive Statistics}
\subsubsection{Distribution of Question Types}
Our question datasets comprises 10,000 questions balanced across domains and reasoning types.  Table \ref{tab:question_features} shows the distribution of questions, which are composed of five different question types (composite, inverse, negation, temporal, and transitive reasoning) across ten knowledge domains. Each cell contains 200 questions, resulting in 1,000 questions per domain and 2,000 questions per question type.
\begin{table}[h!]
\begin{center}
\begin{small}
\caption{Distribution of Question Types Across Domains}
\label{tab:question_features}
\begin{tabular}{lcccccr}
\toprule
\textbf{Domain} & \textbf{Composite} & \textbf{Inverse} & \textbf{Negation} & \textbf{Temporal} & \textbf{Transitive} & \textbf{Total} \\

\midrule
Culture & 200 & 200 & 200 & 200 & 200 & 1,000 \\  \addlinespace
Geography & 200 & 200 & 200 & 200 & 200 & 1,000 \\  \addlinespace
Health & 200 & 200 & 200 & 200 & 200 & 1,000 \\  \addlinespace
History & 200 & 200 & 200 & 200 & 200 & 1,000 \\  \addlinespace
Math & 200 & 200 & 200 & 200 & 200 & 1,000 \\  \addlinespace
Nature & 200 & 200 & 200 & 200 & 200 & 1,000 \\  \addlinespace
People & 200 & 200 & 200 & 200 & 200 & 1,000 \\  \addlinespace
Religion & 200 & 200 & 200 & 200 & 200 & 1,000 \\  \addlinespace
Society & 200 & 200 & 200 & 200 & 200 & 1,000 \\  \addlinespace
Technology & 200 & 200 & 200 & 200 & 200 & 1,000 \\  \addlinespace
\midrule
\textbf{Total} & 2,000 & 2,000 & 2,000 & 2,000 & 2,000 & 10,000 \\
\bottomrule
\end{tabular}
\end{small}
\end{center}

\end{table}

\subsubsection{Distribution of Confidence Measures}
\begin{figure}[!htbp]
    \centering
\includegraphics[width=1.1\textwidth]{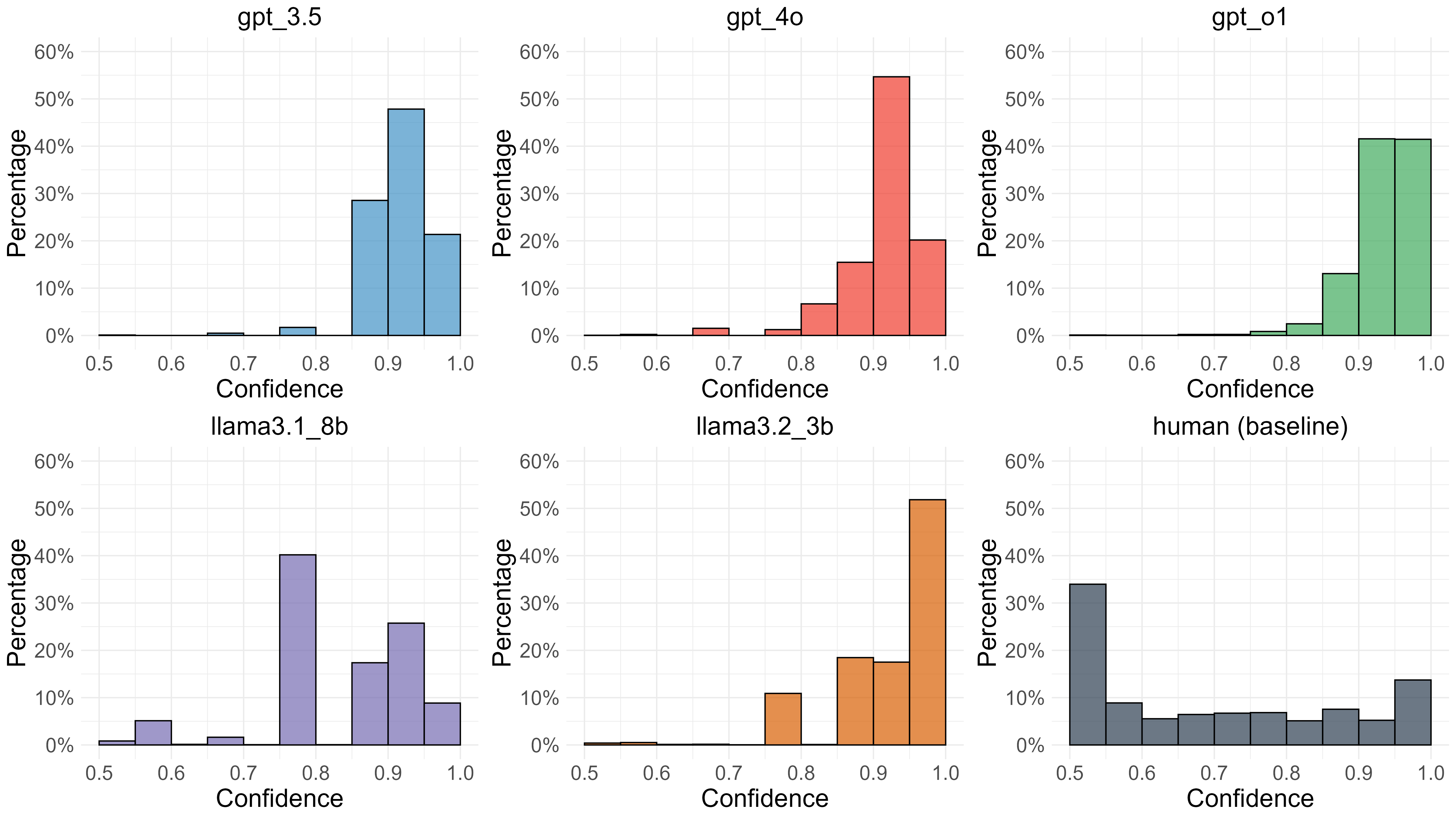}
    \caption{Distribution of Confidence in Answer Correctness across LLMs}
    \label{fig:hist_conf_answer}
\end{figure}

\begin{figure}
    \centering
    \includegraphics[width=1.1\textwidth]{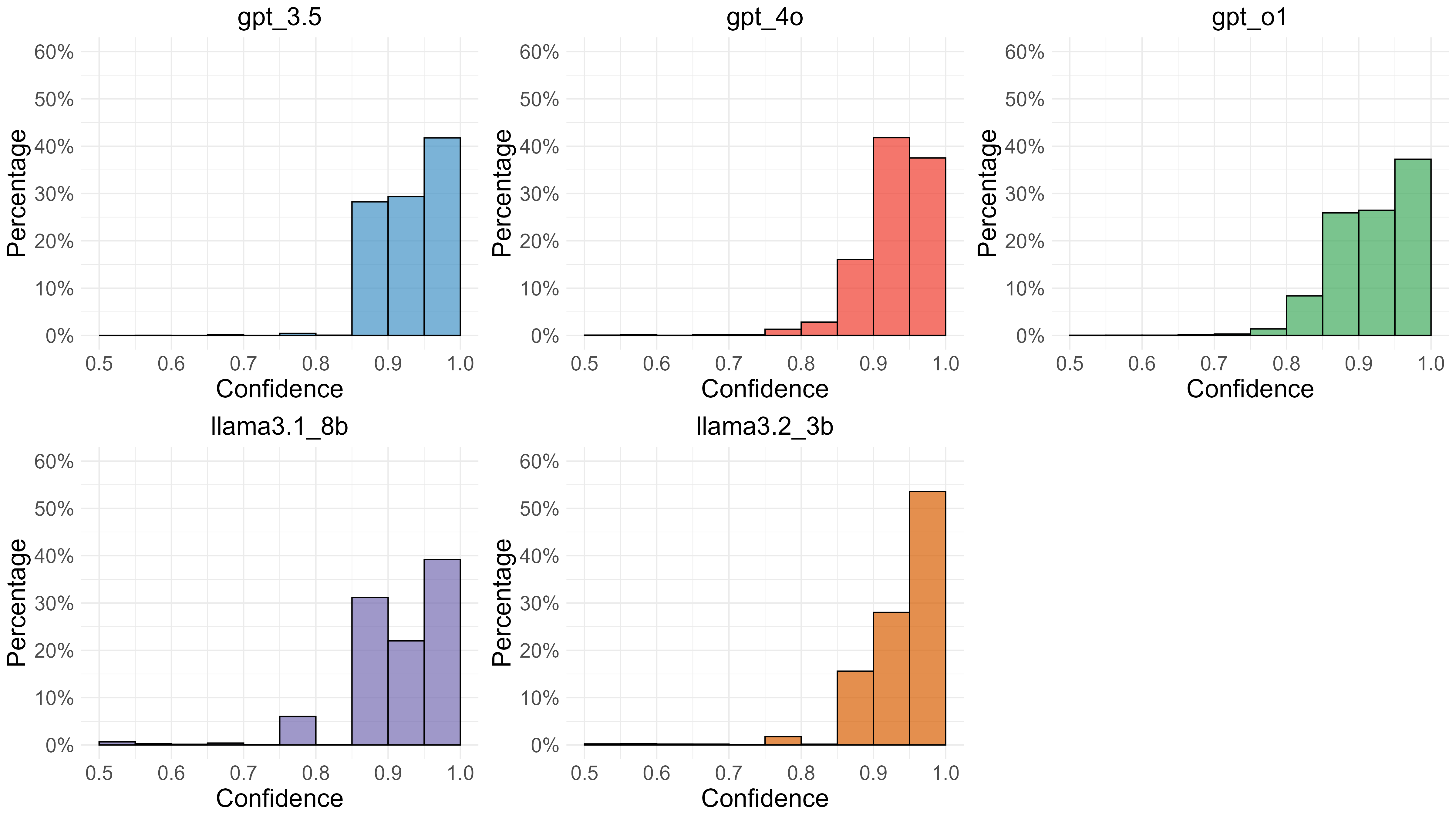}
    \caption{Distribution of Confidence in Facts Correctness across LLMs}
    \label{fig:hist_conf_facts}
\end{figure}

\begin{figure}
    \centering
    \includegraphics[width=1.1\textwidth]{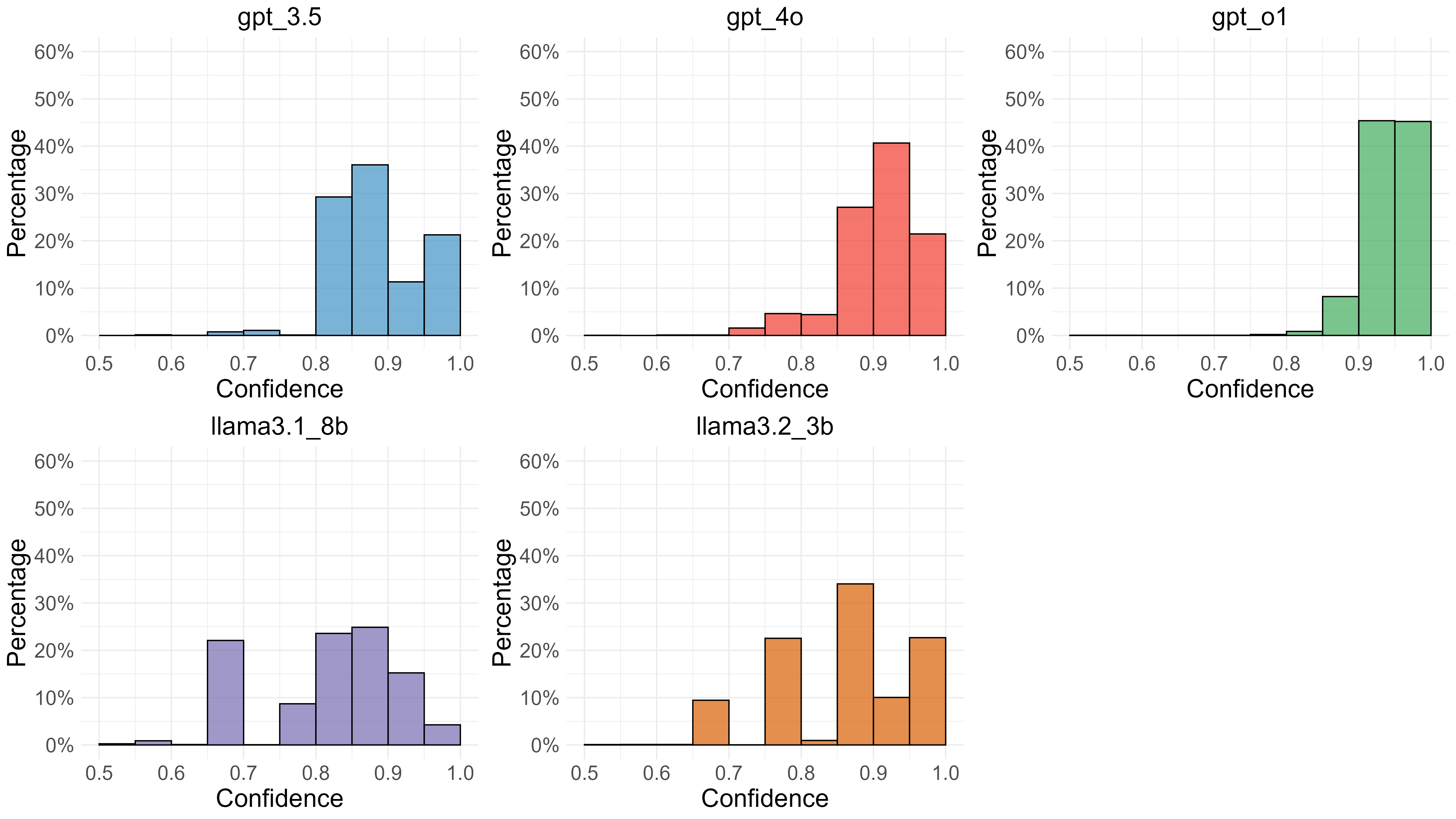}
    \caption{Distribution of Confidence in Reasoning Correctness across LLMs}
    \label{fig:hist_conf_reasoning}
\end{figure}

\begin{figure}
    \centering
\includegraphics[width=1.1\textwidth]{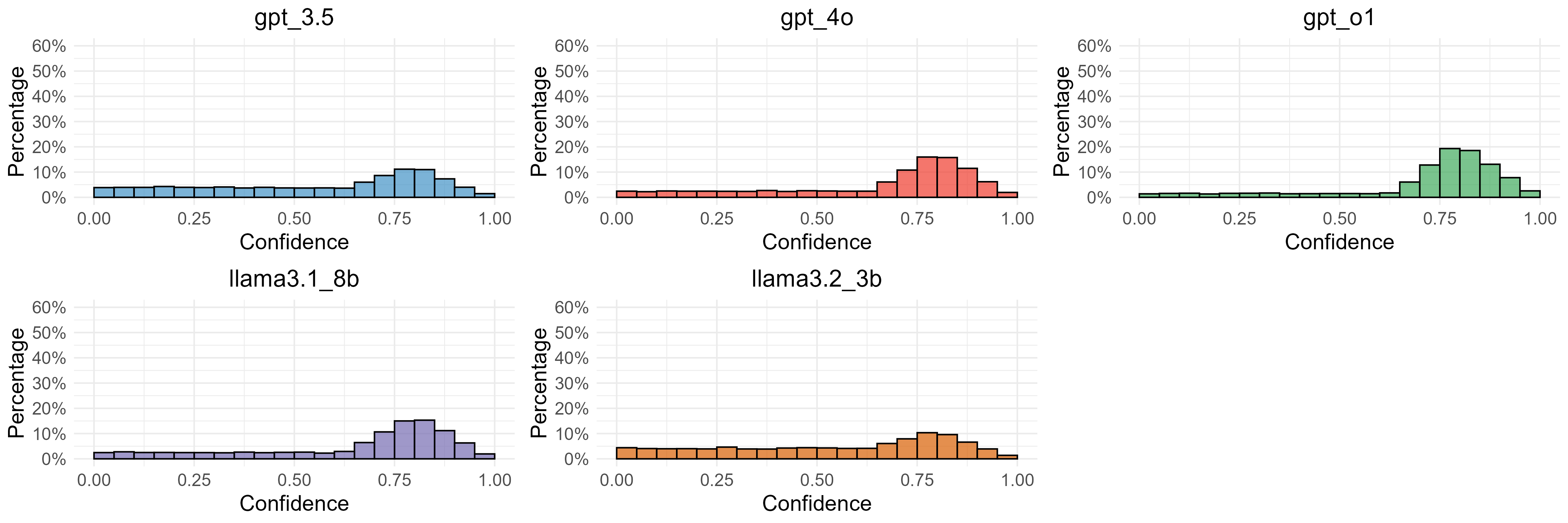}
    \caption{Distribution of Derived Confidence in Correctness across LLMs}
    \label{fig:hist_derive_conf}
\end{figure}

The figures below show the distribution of different confidence measures across all five LLM models in our analysis. In all figures, the x-axis represents confidence levels, and the y-axis shows the percentage of observations in each bin with the width of 0.05. Figure \ref{fig:hist_conf_answer} displays the distribution of self-reported confidence in answer correctness; Figure \ref{fig:hist_conf_facts} and Figure \ref{fig:hist_conf_reasoning} show the distribution of confidence in factual knowledge and reasoning process, respectively; Figure \ref{fig:hist_derive_conf} presents the distribution of our derived confidence measure (multiplication of confidence in facts and reasoning, with the range from 0 to 1). 
 
The distributions reveal several notable patterns. First, all models exhibit a strong rightward skew in their confidence distributions, with a substantial mass of observations at very high confidence levels. This pattern is particularly pronounced for the GPT family of models. Second, the derived confidence measure (Figure \ref{fig:hist_derive_conf}) shows generally lower and more dispersed values compared to the direct self-reported confidence, consistent with our conjunction fallacy hypothesis. Third, there are systematic differences between model families, with the Llama models showing somewhat more dispersed confidence distributions compared to their GPT counterparts.


\clearpage

\clearpage

\subsection{Similarity Measure Results}

Table \ref{tab:corr_gpt_models} and Table \ref{tab:corr_llama_models} present pairwise correlations between the main variables for GPT and llama family of models, respectively. Here, we consider two types of similarity scores: similarity scores in average and similarity score in maximum, both computed for reasoning and facts components separately. There are high correlations between average and maximum scores scross all models, suggesting that these alternative measures could capture the similar underlining patterns. 

In both tables, the positive relationship between self-reported (derived) confidence and similarity measures show consistent patterns for both facts and reasoning components. The correlations coefficients between relf-reported (derived) confidence and similarity are highly significant across all the model, though the magnitudes vary between two families of models.

\begin{table}[htbp]
\centering
\caption{Correlation Matrices: GPT Models}
\label{tab:corr_gpt_models}
\begin{tabular}{lccccccccc}
\multicolumn{10}{c}{\textbf{Panel A: GPT 3.5}} \\
\resizebox{\textwidth}{!}{
\begin{tabular}{lccccccccc}
\hline
Variables &  (1) & (2) & (3) & (4) & (5) & (6) & (7) & (8) & (9) \\ \hline
1. Correct Answer & 1 &  &  &  &  &  &  &  & \\\addlinespace
2. Confidence in Answer &  0.12*** & 1 &  &  &  &  &  &  & \\\addlinespace
3. Derived Confidence &  0.06*** &  0.88*** & 1 &  &  &  &  &  & \\\addlinespace
4. Confidence in Facts & -0.02    &  0.49*** &  0.76*** & 1 &  &  &  &  & \\\addlinespace
5. Confidence in Reasoning &  0.09*** &  0.92*** &  0.89*** &  0.39*** & 1 &  &  &  & \\\addlinespace
6. Average Fact Score &  0.04*** &  0.17*** &  0.20*** &  0.17*** &  0.16*** & 1 &  &  & \\\addlinespace
7. Average Reasoning Score & -0.01    &  0.15*** &  0.17*** &  0.15*** &  0.14*** &  0.89*** & 1 &  & \\\addlinespace
8. Maximum Fact Score &  0.04*** &  0.18*** &  0.20*** &  0.16*** &  0.16*** &  0.98*** &  0.87*** & 1 & \\\addlinespace
9. Maximum Reasoning Score & -0.01    &  0.15*** &  0.17*** &  0.14*** &  0.14*** &  0.87*** &  0.98*** &  0.89*** & 1\\\addlinespace
\hline
\end{tabular}
}
 \\
\addlinespace
\multicolumn{10}{c}{\textbf{Panel B: GPT 4o}} \\
\resizebox{\textwidth}{!}{
\begin{tabular}{lccccccccc}
\hline
Variables &  (1) & (2) & (3) & (4) & (5) & (6) & (7) & (8) & (9) \\ \hline
1. Correct Answer & 1 &  &  &  &  &  &  &  & \\\addlinespace
2. Confidence in Answer & 0.27*** & 1 &  &  &  &  &  &  & \\\addlinespace
3. Derived Confidence & 0.26*** & 0.88*** & 1 &  &  &  &  &  & \\\addlinespace
4. Confidence in Facts & 0.21*** & 0.76*** & 0.91*** & 1 &  &  &  &  & \\\addlinespace
5. Confidence in Reasoning & 0.26*** & 0.89*** & 0.95*** & 0.74*** & 1 &  &  &  & \\\addlinespace
6. Average Fact Score & 0.10*** & 0.24*** & 0.25*** & 0.18*** & 0.27*** & 1 &  &  & \\\addlinespace
7. Average Reasoning Score & 0.07*** & 0.23*** & 0.23*** & 0.17*** & 0.25*** & 0.91*** & 1 &  & \\\addlinespace
8. Maximum Fact Score & 0.10*** & 0.24*** & 0.24*** & 0.18*** & 0.25*** & 0.96*** & 0.88*** & 1 & \\\addlinespace
9. Maximum Reasoning Score & 0.07*** & 0.23*** & 0.23*** & 0.17*** & 0.24*** & 0.87*** & 0.96*** & 0.90*** & 1\\\addlinespace
\hline
\end{tabular}
}
 \\
\addlinespace
\multicolumn{10}{c}{\textbf{Panel C: GPT o1}} \\
\resizebox{\textwidth}{!}{
\begin{tabular}{lccccccccc}
\hline
Variables &  (1) & (2) & (3) & (4) & (5) & (6) & (7) & (8) & (9) \\ \hline
1. Correct Answer & 1 &  &  &  &  &  &  &  & \\\addlinespace
2. Confidence in Answer & 0.24*** & 1 &  &  &  &  &  &  & \\\addlinespace
3. Derived Confidence & 0.26*** & 0.87*** & 1 &  &  &  &  &  & \\\addlinespace
4. Confidence in Facts & 0.23*** & 0.81*** & 0.97*** & 1 &  &  &  &  & \\\addlinespace
5. Confidence in Reasoning & 0.25*** & 0.84*** & 0.89*** & 0.75*** & 1 &  &  &  & \\\addlinespace
6. Average Fact Score & 0.12*** & 0.13*** & 0.13*** & 0.10*** & 0.16*** & 1 &  &  & \\\addlinespace
7. Average Reasoning Score & 0.10*** & 0.13*** & 0.13*** & 0.10*** & 0.16*** & 0.90*** & 1 &  & \\\addlinespace
8. Maximum Fact Score & 0.11*** & 0.14*** & 0.13*** & 0.10*** & 0.16*** & 0.97*** & 0.87*** & 1 & \\\addlinespace
9. Maximum Reasoning Score & 0.09*** & 0.13*** & 0.13*** & 0.10*** & 0.15*** & 0.87*** & 0.97*** & 0.90*** & 1\\\addlinespace
\hline
\end{tabular}
}

\end{tabular}
\begin{tablenotes}
\small
\item \textit{Notes:} $^{***}$p $<$ 0.01; $^{**}$p $<$ 0.05; $^{*}$p $<$ 0.1.
\end{tablenotes}
\end{table}

\begin{table}[htbp]
\centering
\caption{Correlation Matrices: Llama Models}
\label{tab:corr_llama_models}
\begin{tabular}{lccccccccc}
\multicolumn{10}{c}{\textbf{Panel A:Llama 3.1}} \\
\resizebox{\textwidth}{!}{
\begin{tabular}{lccccccccc}
\hline
Variables &  (1) & (2) & (3) & (4) & (5) & (6) & (7) & (8) & (9) \\ \hline
1. Correct Answer & 1 &  &  &  &  &  &  &  & \\\addlinespace
2. Confidence in Answer & 0.28*** & 1 &  &  &  &  &  &  & \\\addlinespace
3. Derived Confidence & 0.22*** & 0.77*** & 1 &  &  &  &  &  & \\\addlinespace
4. Confidence in Facts & 0.18*** & 0.77*** & 0.85*** & 1 &  &  &  &  & \\\addlinespace
5. Confidence in Reasoning & 0.20*** & 0.67*** & 0.95*** & 0.65*** & 1 &  &  &  & \\\addlinespace
6. Average Fact Score & 0.06*** & 0.17*** & 0.14*** & 0.14*** & 0.13*** & 1 &  &  & \\\addlinespace
7. Average Reasoning Score & 0.05*** & 0.17*** & 0.14*** & 0.14*** & 0.13*** & 0.90*** & 1 &  & \\\addlinespace
8. Maximum Fact Score & 0.06*** & 0.20*** & 0.16*** & 0.16*** & 0.15*** & 0.96*** & 0.87*** & 1 & \\\addlinespace
9. Maximum Reasoning Score & 0.05*** & 0.20*** & 0.16*** & 0.16*** & 0.14*** & 0.87*** & 0.96*** & 0.90*** & 1\\\addlinespace
\hline
\end{tabular}
}
 \\
\addlinespace
\multicolumn{10}{c}{\textbf{Panel B: Llama 3.2}} \\
\resizebox{\textwidth}{!}{
\begin{tabular}{lccccccccc}
\hline
Variables &  (1) & (2) & (3) & (4) & (5) & (6) & (7) & (8) & (9) \\ \hline
1. Correct Answer & 1 &  &  &  &  &  &  &  & \\\addlinespace
2. Confidence in Answer & 0.09*** & 1 &  &  &  &  &  &  & \\\addlinespace
3. Derived Confidence & 0.14*** & 0.73*** & 1 &  &  &  &  &  & \\\addlinespace
4. Confidence in Facts & 0.09*** & 0.53*** & 0.78*** & 1 &  &  &  &  & \\\addlinespace
5. Confidence in Reasoning & 0.13*** & 0.72*** & 0.94*** & 0.52*** & 1 &  &  &  & \\\addlinespace
6. Average Fact Score & 0.03**  & 0.16*** & 0.14*** & 0.09*** & 0.15*** & 1 &  &  & \\\addlinespace
7. Average Reasoning Score & 0.03**  & 0.16*** & 0.14*** & 0.09*** & 0.14*** & 0.90*** & 1 &  & \\\addlinespace
8. Maximum Fact Score & 0.04*** & 0.16*** & 0.14*** & 0.10*** & 0.14*** & 0.97*** & 0.87*** & 1 & \\\addlinespace
9. Maximum Reasoning Score & 0.03*** & 0.16*** & 0.14*** & 0.10*** & 0.14*** & 0.88*** & 0.97*** & 0.90*** & 1\\\addlinespace
\hline
\end{tabular}
}
 
\end{tabular}
\begin{tablenotes}
\small
\item \textit{Notes:} $^{***}$p $<$ 0.001; $^{**}$p $<$ 0.01; $^{*}$p $<$ 0.05.
\end{tablenotes}
\end{table}

\section{The welfare effects of LLM exposure} \label{sec:model}

This section describes the welfare analysis discussed in the main body of the paper.

\subsection{The setup}
Assume there is an investment project. The payoff from the project depend on (i) choosing the true state of the world (labeled $Y$ or $N$, in analogy to the questions we ask in our experiment), and (ii) an investment $e$ that increases the payoff from the project if the correct answer is chosen. Investment has a cost $c(e)$, with $c'(e)>0$ and $c''(e) \ge 0$.
If the correct answer is chosen, the payoff is $\pi e$. If the incorrect answer is chosen, the payoff is zero. 

The has beliefs $\widetilde{p} = p + b$, where $p$ is the true probability that the correct state is chosen, and $b$ is an individual bias in assessing the accuracy of the judgment. We focus here on the case where $b>0$, i.e. the individual is overconfident in his response. 

The individual thus chooses the state of the world that has $\widetilde{p} \ge 0.5$. His perceived utility from investing in the project is given by
\begin{align} \label{eq:obj_V}
V(e; \widetilde{p}) = (p+b) \cdot \pi \cdot e - c(e) 
\end{align}

The individual maximizes equation \eqref{eq:obj_V} by choosing effort such that
\begin{align} \label{eq:foc_e}
\frac{\partial V}{\partial e} &= (p+b)\pi - c'(e) = 0
\end{align}
i.e. he chooses effort such that the marginal cost of effort $c'(e)$ equal the perceived marginal benefit $\pi (p+b)$. Optimal perceived effort, denoted $e(p+b)$ henceforth, is thus defined by \begin{align} \label{eq:e_opt}
	e(p+b):  c'(e) = (p+b)\pi
\end{align}
It is straightforward to show that $ \frac{\partial e(p+b)}{\partial \widetilde{p}}>0$. We will use the shorthand notation $e'()$ to refer to this derivate below. Notice that an overconfident individual overinests into the project, since he perceives the the marginal benefit to be $\pi(p+b)$ instead of the true $\pi p$. 

We will at times use the special case of $c(e) = \exp(e \gamma)/\gamma$. In this case, $c'(e) = \exp(e\gamma)$. Optimal effort is given by $e(p+b) =  \frac{\ln(\pi(p+b))}{\gamma}$, and $e'(p+b) = \frac{1}{\gamma(p+b)}$. 

Following \citet{bernheim2019behavioral}, the indivdiual's true welfare is given by
\begin{equation} \label{eq:welfare}
W(p,b) = p\pi e(p+b) - c(e(p+b))
\end{equation}
i.e. it evaluates the welfare using the objective probability $p$, but takes into account that the individual chooses effort acording to equation \eqref{eq:e_opt}, based on his overconfident belief $\widetilde{p}$. This allows us to get a first glimpse at the welfare effects of changing $p$ and $b$. 

In this model, exposure to LLMs potentially changes, both, $p$ and $b$. To derive the first two results we discuss in the text, we evaluate the first derivatives of $W$ with regard to these two parameters. 
The derivative of $W$ with respect to $b$ is given by
\begin{align} \label{eq:dWdb}
	\frac{\partial W}{\partial b} &= p\pi e^\prime (p+b) - c'(e) \cdot e^\prime (p+b)\\ \nonumber
	&= -b\pi e' < 0
\end{align}
where the simplications in the second line of the equation follow from observing that the indivdiual chooses $e$ such that $c'(e) = \pi(p+b)$, and substituting this. Unsurprisingly, welfare decreases in $b$ unambiguously, reflecting the fact that a higher $b$ leads to higher over-investment. 

The derivative of $W$ with respect to $p$ is given by 
\begin{align} \label{eq:dWdp}
	\frac{\partial W}{\partial p} &= \pi e + p\pi e^\prime- c'(e)\\ \nonumber
	&= \pi e - b\pi e^\prime 
\end{align}
Notice that, somewhat surprisingly, the derivative is not unambiguously positive if $b>0$ and $e'>0$. That is, it is possible that welfare decreases if the objective proability of success increases. We discuss this effect in more detail below. 

These derivatives allow us to state the first two results. 

\noindent \medskip \emph{\textbf{Result 1:} if $e'(p+b)=0$, then welfare unambiguously increases in $p$.}

This is a very intuitive result. Since $e'(p+b)=0$, effort does not change a function of $\widetilde{p}$. Thus, there is no cost from overinvestment, and payoffs unambiguously increase as $p$ increases.

\noindent \medskip \emph{\textbf{Result 2:} If $e'(p+b)>0$, then welfare decreases in $p$ if \begin{align*} 
		\pi e < b\pi e^\prime 
\end{align*} In the case of an exponential cost function, this is the case if \begin{align} \label{eq:result2}
\ln(\pi(p+b)) < \frac{b}{p+b}
\end{align}}
Intuitively speaking, the condition for this effect to occur is if $e$ is relatively low, but $e(p+b)$ is elastic. If $e(p+b)$ responds very strong to changes in $p$. In this case, an increase in $p$ has a direct on utility, raising payoffs by $\pi e$ per unit of $p$. However, a larger $p$ also increases investment. Recall however, that, the individual over-invests because of $b$. Thus, this increase creates increasingly expensive over-investment. If effort is elastic enough, this second effect can dominate the first and decrease overall welfare. 

Equation \eqref{eq:result2} spells out the condition for the case of an exponential cost function. The left-hand side of equation \eqref{eq:result2} needs to be positive for $e$ to be positive. The condition for effort to be decreasing in $p$ is that $\ln(\pi(p+b)) $ is bounded from above by $b/(p+b)$, which can be interpreted as the percent of the subjective belief in success due to overconfidence. 
\footnote{In the case of an exponential cost function, it is not necessary for effort to be low to generate a negative effect of $p$ on $W$. If $\gamma$ is low, e can be large even if $\ln(\pi(p+b)) $ is only slightly positive.}

\subsection{The welfare effects of changes in $p$ and $b$}
We model exposure to LLM input as a simulatneous change in $p$ and $b$. We now develop the Taylor approximation to $W$ to characterize the approximate effect of discrete changes in $p$ and $b$ on welfare. 

As is common in these approximations, we assume that $e''(p+b) \approx 0$ and that these terms can therefore be neglected. We then calcuate the second-order approximation from the remaining terms to obtain  \begin{align} \label{eq:taylor_gen}
W(p+\Delta p + b + \Delta b) -W(p,b) &\approx \  \pi \big(    \underbrace{\left( e - b e'\right)\Delta p + \frac{e^\prime}{2} \cdot (\Delta p)^2}_{\text{ambiguous}} \\ \nonumber
&\underbrace{- b \cdot e^\prime \Delta b + \frac{- e^\prime}{2} \cdot (\Delta b)^2}_{\text{negative}} \big)
\end{align}

In order to evaluate a particular change in, both, $p$ and $b$, we need to make assumptions about their relative magnitude. Motivated in part by the empirical evidence from our experiment, we assume that $\Delta p = \Delta b>0$ due to exposure to LLM input.

\noindent
\medskip
\textbf{\emph{Result 3.}} Assume that $\Delta b = \alpha \Delta p$ with $\alpha \ge 1$. The combined change reduces welfare if:
\begin{align*}
    \pi e < (1+\alpha) b\pi e^\prime
\end{align*}
In the case of an exponential cost function, this is the case if:
\begin{align*} \label{eq:result2}
    \ln(\pi(p+b)) < (1+\alpha)\frac{b}{p+b}
\end{align*}

The intuition behind Result 3 is most easily see for the case where $\Delta p $ and $\Delta b$ are approximately the same. The conditions for LLM exposure to increase welfare become even stricter than in Result 2. The reason is that, in addition to the dampening effect on the welfare gains from $p$, now the direct negative effects keep piling on. The case where the two effects are of the same magnitude is particularly simple, because the the second-order effects cancel. 

If, more generally, $\Delta b = \alpha \Delta p$, where $\alpha >1$, the inequality would be even easier to satisfy, because the negative second-order effects from $\Delta b$ outweigh the positive ones from $\Delta p$, as can  be seen in equation \eqref{eq:taylor_gen}.

The final result provides a characterization of the welfare loss due to bias, i.e. it uses the Taylor approximation to calculate by how much welfare increased if $b=0$. This is not explicitly discussed in the text, but we report the quantity here for completeness. 

\noindent \medskip  \emph{\textbf{Result 4:} The welfare gains if bias were removed are given by \begin{align*}
		\Delta W^* = \pi \frac{b^2e'}{2}
\end{align*}}

\subsection{Piecing together the second derivatives for the Taylor approximation}
As is customary in derive these approximations  , we assume that behavior effects beyond the first order can be ignored, i.e. that $e ^{\prime \prime} (p +b) = 0$. 
Using this, we obtain
\begin{align*}
\frac{\partial ^2 W}{\partial b^2} &= -\pi e^2 \\
\frac{\partial ^2 W}{\partial p^2} &= \pi e^2 \\
\end{align*}

For the cross-derivative, we obtain \begin{align*}
	\frac{\partial}{\partial b}\left(\frac{\partial W}{\partial p}\right) = \frac{\partial ^2 W}{\partial b \partial p}  = \pi \frac{\partial e}{\partial b} - \pi \frac{\partial e}{\partial b} = 0. 
\end{align*}

With the first derivatives defined earlier, we can then piece together the second-order Taylor approximation 
\begin{align*}
	W(p+\Delta p + b + \Delta b) &\approx W(p+b) + \frac{\partial W}{\partial p} \cdot \Delta p + \frac{\partial^2 W}{(\partial p)^2} \cdot \frac{1}{2} \cdot (\Delta p)^2 \\
	&+ \frac{\partial W}{\partial b} \cdot \Delta b + \frac{\partial^2 W}{(\partial b)^2} \cdot \frac{1}{2} (\Delta b)^2 \\
	&= W(p+b) + \left(\pi e - b\pi e'\right)\Delta p + \frac{\pi e^\prime}{2} \cdot (\Delta p)^2 \\
	&- b\pi \cdot e^\prime \Delta b + \frac{-\pi e^\prime}{2} \cdot (\Delta b)^2
\end{align*}

\section{Robustness Checks}
\subsection{Change in Prompt}

We examined two variants to the baseline prompts: in the No-Frame prompt, instead of asking for a yes/no answer and then eliciting the confidence in the answer, we asked directly "what is the probability, that the correct answer is `No'?" Similary, in the Yes-Frame prompt, we asked "what is the probability, that the correct answer is `Yes'?" From these statements, we a yes/no answer as the event with more than 50\% probability. This allows us to establish analogous measures of accuracy and confidence in the answer. Table \ref{tab:prompt frames} displays the results. There is no systematic or quantitatively important influence of the alternative framing on accuracy or confidence.

\begin{table}
\centering
\caption{Model Performance Across Different Prompting Frames}
\label{tab:prompt frames}
\begin{tabular}{lccccccc}
\toprule
& \multicolumn{2}{c}{\textbf{Baseline Prompt}} & \multicolumn{2}{c}{\textbf{No-Frame Prompt}} & \multicolumn{2}{c}{\textbf{Yes-Frame Prompt}} \\
\cmidrule(lr){2-3} \cmidrule(lr){4-5} \cmidrule(lr){6-7}
\textbf{Model} & Accuracy & Confidence & Accuracy & Confidence & Accuracy & Confidence \\
\midrule
GPT-3.5 & 0.348 & 0.942 & 0.361 & 0.945 & 0.351 & 0.906 \\ \addlinespace
GPT-4o & 0.635 & 0.937 & 0.567 & 0.964 & 0.651 & 0.943 \\ \addlinespace
GPT-o1 & 0.735 & 0.954 & 0.706 & 0.973 & 0.744 & 0.958 \\ \addlinespace
Llama 3.1 (8B) & 0.626 & 0.858 & 0.630 & 0.981 & 0.533 & 0.945 \\ \addlinespace
Llama 3.2 (3B) & 0.615 & 0.944 & 0.638 & 0.991 & 0.425 & 0.991 \\ \addlinespace
\bottomrule
\end{tabular}
\end{table}

\subsection{Change in Temperature} \label{sec:check temperature}

We also examine whether our findings are robust to variations in the temperature parameter, which controls the randomness of model outputs. Higher temperature settings increase output variability and may affect both accuracy and confidence assessments. We test temperatures of 0, 0.6, and 1.0 for all models except GPT-o1, which does not support temperature adjustment.

Table \ref{tab:summary_temp_performance} presents the results across different temperature settings. The patterns we observe are consistent across temperature variations: accuracy rates remain relatively stable, with only modest changes (typically within 2-3 percentage points) as temperature increases. Similarly, confidence levels show minimal sensitivity to temperature adjustments, varying by less than 2 percentage points in most cases. For Llama models, slightly higher temperatures appear to marginally improve accuracy while reducing confidence, but these effects are small in magnitude.

These results demonstrate that our core findings regarding LLM overconfidence are robust to different temperature settings. The systematic bias we document persists regardless of the degree of randomness in model outputs, suggesting that overconfidence is a fundamental characteristic of these models rather than an artifact of specific parameter configurations.
\begin{table}[htbp]
\centering
\caption{Model Performance at Different Temperatures}
\label{tab:summary_temp_performance}
\begin{tabular}{lccccccc}
\toprule
& \multicolumn{2}{c}{\textbf{Temp = 0}} & \multicolumn{2}{c}{\textbf{Temp = 0.6}} & \multicolumn{2}{c}{\textbf{Temp = 1}} \\
\cmidrule(lr){2-3} \cmidrule(lr){4-5} \cmidrule(lr){6-7}
\textbf{Model} & \textbf{Accuracy} & \textbf{Confidence} & \textbf{Accuracy} & \textbf{Confidence} & \textbf{Accuracy} & \textbf{Confidence} \\
\midrule
gpt\_3.5 & 0.348 & 0.942 & 0.362 & 0.939 & 0.374 & 0.934 \\
gpt\_4o & 0.635 & 0.937 & 0.646 & 0.931 & 0.645 & 0.927 \\
gpt\_o1 & 0.735 & 0.954 & -- & -- & -- & -- \\
llama3.1 (8b) & 0.626 & 0.858 & 0.651 & 0.829 & 0.657 & 0.820 \\
llama3.2 (3b) & 0.615 & 0.944 & 0.545 & 0.943 & 0.529 & 0.926 \\
\bottomrule
\end{tabular}
\end{table}

We provide detailed regression analyses examining how temperature settings affect model performance across different prompting frames. The baseline prompt results (Table \ref{tab:robustness_baseline_temp_regression}) reveal systematic temperature effects across all models. Higher temperatures generally lead to modest accuracy improvements for GPT models and Llama 3.1, while Llama 3.2 shows deteriorating performance. Confidence levels consistently decline with increasing temperature, creating interesting dynamics in bias patterns. For most models, the combined effect results in reduced overconfidence at higher temperatures, though substantial bias persists even at elevated temperature settings.

The alternative prompting frames (Yes-frame and No-frame, Table \ref{tab:robustness_yesframe_temp_regression} and \ref{tab:robustness_noframe_temp_regression} respectively) demonstrate the robustness of our core findings while revealing some frame-specific patterns. In the Yes-frame condition, where models assess the probability that "Yes" is correct, we observe different baseline confidence levels but similar temperature gradients. The No-frame results, where models evaluate the probability that "No" is correct, show comparable patterns with slight variations in magnitude. Specifically, there is  substantial increase in non-response rates at temperature 1.5, particularly for GPT-4o and Llama models, where missing response rates can exceed 80\%. We address this by reporting both standard accuracy measures (treating missing responses as missing data) and a "CorrectOverall" measure that treats non-responses as incorrect answers, providing a conservative assessment of model performance under extreme temperature settings.

\begin{table}[htbp]
   \caption{Temperature Effects on Model Performance: Baseline Prompt}
   \label{tab:robustness_baseline_temp_regression}
   \centering
   \renewcommand{\arraystretch}{0.9}
   \begin{tabular}{lcccc}
      \toprule
      & \multicolumn{4}{c}{\textbf{Results}} \\
      \textbf{Dependent Variable} & \textbf{GPT 3.5} & \textbf{GPT 4o} & \textbf{Llama 3.1} & \textbf{Llama 3.2} \\
      \midrule
      \multicolumn{5}{l}{\textbf{Panel A: Accuracy}} \\
      \addlinespace
      Baseline (T=0) & 0.3337 & 0.6237 & 0.5949 & 0.7013 \\
      \addlinespace
      T=0.6 & 0.0140** & 0.0110*** & 0.0257*** & -0.0695*** \\
      & (0.0045) & (0.0024) & (0.0047) & (0.0049) \\
      T=1 & 0.0281*** & 0.0103*** & 0.0314*** & -0.0862*** \\
      & (0.0050) & (0.0026) & (0.0051) & (0.0055) \\
      T=1.5 & 0.0825*** & 0.0101** & 0.0104 & -0.0870*** \\
      & (0.0054) & (0.0032) & (0.0122) & (0.0073) \\
      \addlinespace
      \midrule
      \multicolumn{5}{l}{\textbf{Panel B: Confidence}} \\
      \addlinespace
      Baseline (T=0) & 0.9422 & 0.9372 & 0.8582 & 0.9437 \\
      \addlinespace
      T=0.6 & -0.0034*** & -0.0060*** & -0.0286*** & -0.0010 \\
      & (0.0005) & (0.0004) & (0.0011) & (0.0009) \\
      T=1 & -0.0086*** & -0.0103*** & -0.0412*** & -0.0180*** \\
      & (0.0006) & (0.0005) & (0.0013) & (0.0011) \\
      T=1.5 & -0.0211*** & -0.1065*** & -0.0975*** & -0.1226*** \\
      & (0.0007) & (0.0041) & (0.0105) & (0.0043) \\
      \addlinespace
      \midrule
      \multicolumn{5}{l}{\textbf{Panel C: Bias}} \\
      \addlinespace
      Baseline (T=0) & 0.5941 & 0.3020 & 0.2328 & 0.3286 \\
      \addlinespace
      T=0.6 & -0.0173*** & -0.0170*** & -0.0546*** & 0.0697*** \\
      & (0.0045) & (0.0025) & (0.0047) & (0.0050) \\
      T=1 & -0.0367*** & -0.0205*** & -0.0772*** & 0.0713*** \\
      & (0.0049) & (0.0026) & (0.0053) & (0.0057) \\
      T=1.5 & -0.1043*** & -0.1139*** & -0.1152*** & 0.0346*\\
      & (0.0055) & (0.0081) & (0.0442) & (0.0155) \\
      \addlinespace
      \midrule
      \multicolumn{5}{l}{\textbf{Panel D: CorrectOverall (Punish Non-Answer)}} \\
      \addlinespace
      T=0 & 0.3406 & 0.6347 & 0.6165 & 0.6122 \\
      T=0.6 & 0.3513 & 0.6457 & 0.6190 & 0.5418 \\
      T=1 & 0.3547 & 0.6448 & 0.5938 & 0.5168 \\
      T=1.5 & 0.4144 & 0.5089 & 0.0827 & 0.2839 \\
      \addlinespace
      \midrule
      \textbf{Fixed Effects} & & & & \\
      qid & yes & yes & yes & yes \\
      \midrule
      \textbf{Missing by Temperature (\%)} & & & & \\
      T=0 & 2.05, 2.05 & 0.00, 0.27 & 1.56, 2.50 & 0.43, 0.90 \\
      T=0.6 & 2.99, 3.00 & 0.00, 0.00 & 4.95, 5.81 & 0.61, 1.00 \\
      T=1 & 5.20, 5.20 & 0.09, 0.09 & 9.63, 19.44 & 2.38, 7.23 \\
      T=1.5 & 3.24, 6.39 & 21.01, 88.42 & 87.81, 99.36 & 47.48, 90.40 \\
      \bottomrule
   \end{tabular}
   
\begin{tablenotes}
\item \small \textit{Notes:} Regression analyses use question fixed effects to control for question-specific difficulty,\\ with Temperature = 0 as the reference category. 
Accuracy measures the proportion of correct \\ binary responses (0/1); Confidence represents the model's self-assessed probability of correctness;\\ Bias is calculated as Confidence - Accuracy, where positive values indicate overconfidence. \\
Panel D shows CorrectOverall values, which treat missing responses as incorrect answers.\\ Standard errors in parentheses, clustered by question ID. \\
***: p < 0.001, **: p < 0.01, *: p < 0.05.
\end{tablenotes}

\end{table}

\begin{table}[htbp]
   \caption{Temperature Effects on Model Performance: Yes-Frame Prompt}
   \label{tab:robustness_yesframe_temp_regression}
   \centering
   \begin{tabular}{lcccc}
      \toprule
      & \multicolumn{4}{c}{\textbf{Yes Frame Results}} \\
      \textbf{Dependent Variable} & \textbf{GPT 3.5} & \textbf{GPT 4o} & \textbf{Llama 3.1} & \textbf{Llama 3.2} \\
      \midrule
      \multicolumn{5}{l}{\textbf{Panel A: Accuracy}} \\
      \addlinespace
      Baseline (T=0) & 0.4935 & 0.6739 & 0.5604 & 0.4310 \\
      \addlinespace
      T=0.6 & -0.0100* & -0.0019 & 0.0079 & 0.0267*** \\
      & (0.0047) & (0.0021) & (0.0048) & (0.0056) \\
      T=1 & -0.0220*** & -0.0068** & 0.0123** & 0.0525*** \\
      & (0.0052) & (0.0025) & (0.0053) & (0.0060) \\
      T=1.5 & -0.0471*** & -0.0248*** & 0.0419 & 0.0666*** \\
      & (0.0058) & (0.0031) & (0.0221) & (0.0094) \\
      \addlinespace
      \midrule
      \multicolumn{5}{l}{\textbf{Panel B: Confidence}} \\
      \addlinespace
      Baseline (T=0) & 0.9063 & 0.9434 & 0.9451 & 0.9912 \\
      \addlinespace
      T=0.6 & 0.0099*** & -0.0011 & -0.0143*** & -0.0056*** \\
      & (0.0018) & (0.0007) & (0.0014) & (0.0008) \\
      T=1 & 0.0083*** & -0.0034*** & -0.0256*** & -0.0237*** \\
      & (0.0021) & (0.0008) & (0.0015) & (0.0011) \\
      T=1.5 & 0.0044 & -0.0059*** & -0.0167*** & -0.0501*** \\
      & (0.0022) & (0.0010) & (0.0061) & (0.0022) \\
      \addlinespace
      \midrule
      \multicolumn{5}{l}{\textbf{Panel C: Bias}} \\
      \addlinespace
      Baseline (T=0) & 0.4127 & 0.2695 & 0.3847 & 0.5601 \\
      \addlinespace
      T=0.6 & 0.0198*** & 0.0008 & -0.0222*** & -0.0323*** \\
      & (0.0057) & (0.0025) & (0.0053) & (0.0058) \\
      T=1 & 0.0304*** & 0.0034 & -0.0379*** & -0.0762*** \\
      & (0.0063) & (0.0029) & (0.0059) & (0.0063) \\
      T=1.5 & 0.0515*** & 0.0190*** & -0.0586* & -0.1167*** \\
      & (0.0070) & (0.0036) & (0.0236) & (0.0100) \\
      \addlinespace
      \midrule
      \textbf{Panel D: CorrectOverall (Punish Non-Answer)} & & & & \\
      \addlinespace
      T=0 & 0.4932 & 0.6739 & 0.5552 & 0.4109 \\
      T=0.6 & 0.4823 & 0.6720 & 0.5611 & 0.4327 \\
      T=1 & 0.4677 & 0.6671 & 0.5222 & 0.4362 \\
      T=1.5 & 0.4069 & 0.5873 & 0.0285 & 0.1638 \\
      \addlinespace
      \midrule
      \textbf{Fixed Effects} & & & & \\
      qid & yes & yes & yes & yes \\
      \midrule
      \textbf{Missing by Temperature (\%)} & & & & \\
      T=0 & 0.07 & 0.00 & 0.93 & 4.67 \\
      T=0.6 & 0.43 & 0.00 & 1.25 & 4.53 \\
      T=1 & 1.16 & 0.02 & 9.41 & 9.30 \\
      T=1.5 & 9.70 & 10.63 & 95.91 & 68.01 \\

      \bottomrule
   \end{tabular}
   
  \begin{tablenotes}
\item \small \textit{Notes:} See baseline prompt table notes for variable definitions and methodology. \\
Standard errors in parentheses, clustered by question ID. \\
***: p < 0.001, **: p < 0.01, *: p < 0.05.
\end{tablenotes}
\end{table}

\begin{table}[htbp]
   \caption{Temperature Effects on Model Performance: No-Frame Prompt}
   \label{tab:robustness_noframe_temp_regression}
   \centering
   \begin{tabular}{lcccc}
      \toprule
      & \multicolumn{4}{c}{\textbf{No Frame Results}} \\
      \textbf{Dependent Variable} & \textbf{GPT 3.5} & \textbf{GPT 4o} & \textbf{Llama 3.1} & \textbf{Llama 3.2} \\
      \midrule
      \multicolumn{5}{l}{\textbf{Panel A: Accuracy}} \\
      \addlinespace
      Baseline (T=0) & 0.3013 & 0.5603 & 0.6172 & 0.6294 \\
      \addlinespace
      T=0.6 & 0.0111* & 0.0021 & -0.0179*** & -0.0184*** \\
      & (0.0046) & (0.0034) & (0.0049) & (0.0058) \\
      T=1 & 0.0211*** & -0.0039 & -0.0475*** & -0.0426*** \\
      & (0.0049) & (0.0037) & (0.0054) & (0.0062) \\
      T=1.5 & 0.0598*** & -0.0283*** & -0.0547 & -0.0107 \\
      & (0.0056) & (0.0049) & (0.0308) & (0.0106) \\
      \addlinespace
      \midrule
      \multicolumn{5}{l}{\textbf{Panel B: Confidence}} \\
      \addlinespace
      Baseline (T=0) & 0.9452 & 0.9642 & 0.9809 & 0.9912 \\
      \addlinespace
      T=0.6 & -0.0011 & -0.0028*** & -0.0073*** & -0.0030*** \\
      & (0.0016) & (0.0005) & (0.0010) & (0.0008) \\
      T=1 & -0.0060*** & -0.0068*** & -0.0297*** & -0.0211*** \\
      & (0.0017) & (0.0006) & (0.0012) & (0.0011) \\
      T=1.5 & -0.0158*** & -0.0176*** & -0.0595*** & -0.0453*** \\
      & (0.0019) & (0.0010) & (0.0098) & (0.0025) \\
      \addlinespace
      \midrule
      \multicolumn{5}{l}{\textbf{Panel C: Bias}} \\
      \addlinespace
      Baseline (T=0) & 0.6439 & 0.4039 & 0.3638 & 0.3618 \\
      \addlinespace
      T=0.6 & -0.0122** & -0.0049 & 0.0106* & 0.0153**\\
      & (0.0046) & (0.0034) & (0.0048) & (0.0058) \\
      T=1 & -0.0271*** & -0.0029 & 0.0178*** & 0.0214*** \\
      & (0.0048) & (0.0036) & (0.0052) & (0.0061) \\
      T=1.5 & -0.0756*** & 0.0107* & -0.0048 & -0.0346*** \\
      & (0.0056) & (0.0048) & (0.0295) & (0.0104) \\
      \addlinespace
      \midrule
      \textbf{Panel D: CorrectOverall (Punish Non-Answer)} & & & & \\
      \addlinespace
      T=0 & 0.3013 & 0.5602 & 0.6147 & 0.6251 \\
      T=0.6 & 0.3124 & 0.5622 & 0.5979 & 0.6071 \\
      T=1 & 0.3222 & 0.5563 & 0.5398 & 0.5734 \\
      T=1.5 & 0.3245 & 0.2642 & 0.0151 & 0.1610 \\
      \addlinespace
      \midrule
      
      \textbf{Fixed Effects} & & & & \\
      qid & yes & yes & yes & yes \\
      \midrule
      \textbf{Missing by Temperature (\%)} & & & & \\
      T=0 & 0.01 & 0.02 & 0.40 & 0.68 \\
      T=0.6 & 0.00 & 0.02 & 0.17 & 0.74 \\
      T=1 & 0.06 & 0.01 & 5.53 & 2.50 \\
      T=1.5 & 10.08 & 45.50 & 97.69 & 75.18 \\
   
      \bottomrule
   \end{tabular}
   
   \begin{tablenotes}
\item \small \textit{Notes:} See baseline prompt table notes for variable definitions and methodology. \\
Standard errors in parentheses, clustered by question ID. \\
 ***: p < 0.001, **: p < 0.01, *: p < 0.05.
\end{tablenotes}
\end{table}

\subsection{Replication Results}\label{subsec:replication_results}
\subsubsection{Descriptive Summary}
To validate the reliability of our LLM-generated responses, we conducted four additional rounds of experiments with GPT-3.5 and GPT-4o in December 2024, followed by another round in February 2025, yielding a total of six experimental rounds for each model. Each round maintained identical prompt structures and problem sets (N = 10,000) while implementing randomized sequences of the question list to control for potential order effects. Table \ref{tab:006_replication_summary} provides a detailed summary of the correctness and confidence measures across all six rounds. The fraction of correct answers shows the accuracy rate, while the confidence measures show the model's self-reported confidence in its answers, facts, and reasoning respectively. The mean confidence measure represents the average of all confidence metrics. 
\begin{table}[!h]
\centering
\caption{Replication Results: Accuracy and Confidence Measures}
\centering
\resizebox{\ifdim\width>\linewidth\linewidth\else\width\fi}{!}{
\begin{tabular}[t]{lccccc}
\toprule
\multicolumn{1}{c}{ } & \multicolumn{1}{c}{a) Fraction of} & \multicolumn{3}{c}{b) LLM confidence in ...} & \multicolumn{1}{c}{c) Derived} \\
\cmidrule(l{3pt}r{3pt}){2-2} \cmidrule(l{3pt}r{3pt}){3-5} \cmidrule(l{3pt}r{3pt}){6-6}
Round & correct answers & answer ($\tilde{p}$) & facts ($p_F$) & reasoning ($p_R$) & confidence ($\tilde{p}_{F,R}$)\\
\midrule
\addlinespace[0.3em]
\multicolumn{6}{l}{\textbf{GPT 3.5}}\\
\hspace{1em}1 & 0.35 & 0.94 & 0.96 & 0.91 & 0.87\\
\hspace{1em} & (0.476) & (0.045) & (0.044) & (0.061) & (0.082)\\
\addlinespace
\hspace{1em}2 & 0.35 & 0.94 & 0.95 & 0.91 & 0.87\\
\hspace{1em} & (0.476) & (0.045) & (0.044) & (0.061) & (0.081)\\
\addlinespace
\hspace{1em}3 & 0.35 & 0.94 & 0.95 & 0.91 & 0.87\\
\hspace{1em} & (0.477) & (0.045) & (0.045) & (0.061) & (0.082)\\
\addlinespace
\hspace{1em}4 & 0.35 & 0.94 & 0.96 & 0.91 & 0.87\\
\hspace{1em} & (0.477) & (0.045) & (0.043) & (0.061) & (0.082)\\
\addlinespace
\hspace{1em}5 & 0.35 & 0.94 & 0.96 & 0.91 & 0.87\\
\hspace{1em} & (0.477) & (0.045) & (0.044) & (0.061) & (0.082)\\
\addlinespace
\hspace{1em}6 & 0.35 & 0.94 & 0.95 & 0.91 & 0.87\\
\hspace{1em} & (0.476) & (0.045) & (0.045) & (0.061) & (0.082)\\
\addlinespace
\addlinespace[0.3em]
\multicolumn{6}{l}{\textbf{GPT 4o}}\\
\hspace{1em}1 & 0.63 & 0.94 & 0.95 & 0.93 & 0.88\\
\hspace{1em} & (0.482) & (0.053) & (0.044) & (0.054) & (0.083)\\
\addlinespace
\hspace{1em}2 & 0.65 & 0.93 & 0.94 & 0.92 & 0.87\\
\hspace{1em} & (0.478) & (0.058) & (0.048) & (0.055) & (0.085)\\
\addlinespace
\hspace{1em}3 & 0.65 & 0.93 & 0.94 & 0.92 & 0.87\\
\hspace{1em} & (0.478) & (0.058) & (0.049) & (0.055) & (0.085)\\
\addlinespace
\hspace{1em}4 & 0.64 & 0.93 & 0.94 & 0.92 & 0.87\\
\hspace{1em} & (0.479) & (0.059) & (0.048) & (0.054) & (0.085)\\
\addlinespace
\hspace{1em}5 & 0.64 & 0.93 & 0.94 & 0.92 & 0.87\\
\hspace{1em} & (0.479) & (0.059) & (0.049) & (0.055) & (0.086)\\
\addlinespace
\hspace{1em}6 & 0.65 & 0.93 & 0.94 & 0.93 & 0.87\\
\hspace{1em} & (0.476) & (0.055) & (0.046) & (0.052) & (0.082)\\
\addlinespace
\bottomrule
\label{tab:006_replication_summary}
\end{tabular}}
\end{table}

\begin{figure}[!htbp]
    \centering
\includegraphics[width=1.1\textwidth]{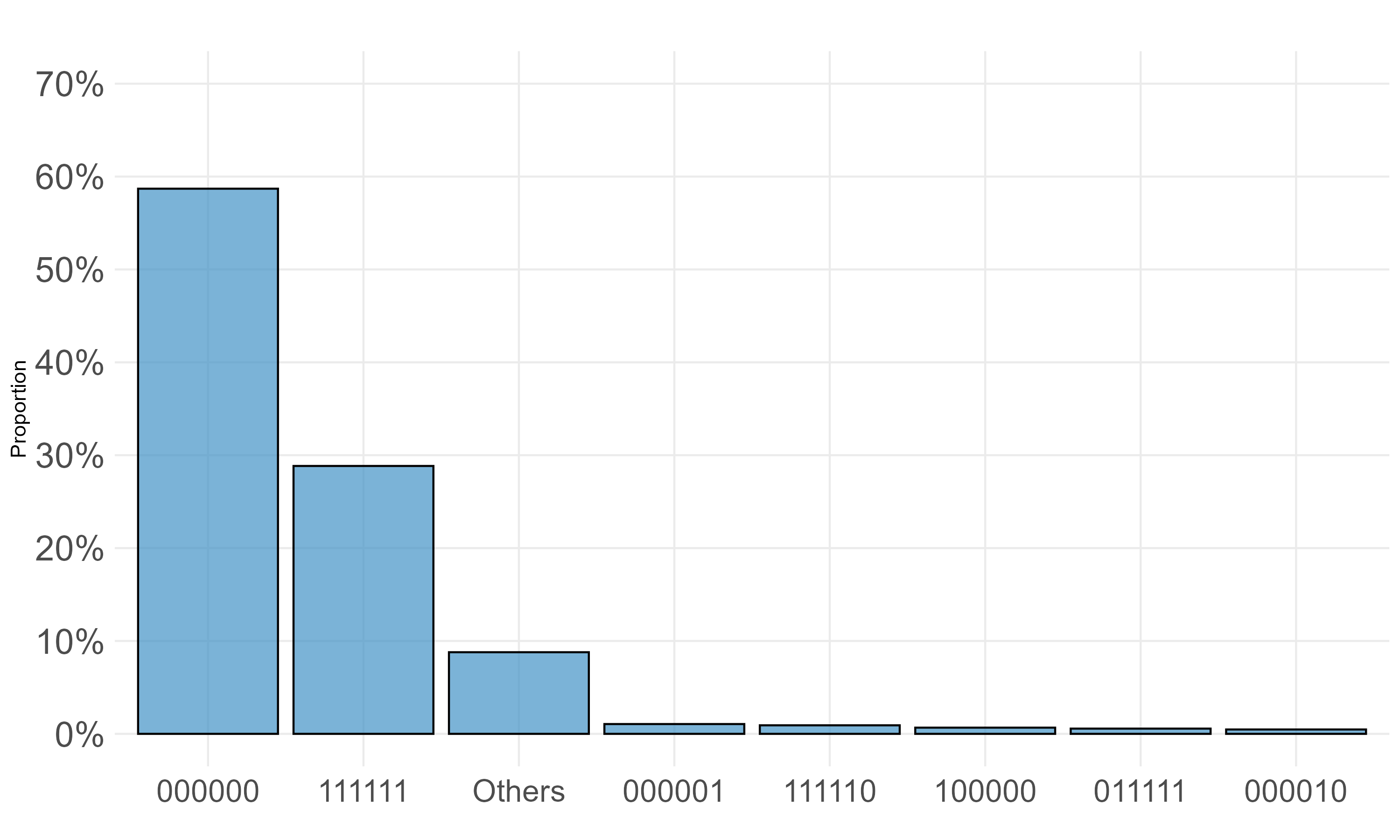}
    \centerline{Panel A: GPT-3.5}
    \vspace{1em}
\includegraphics[width=1.1\textwidth]{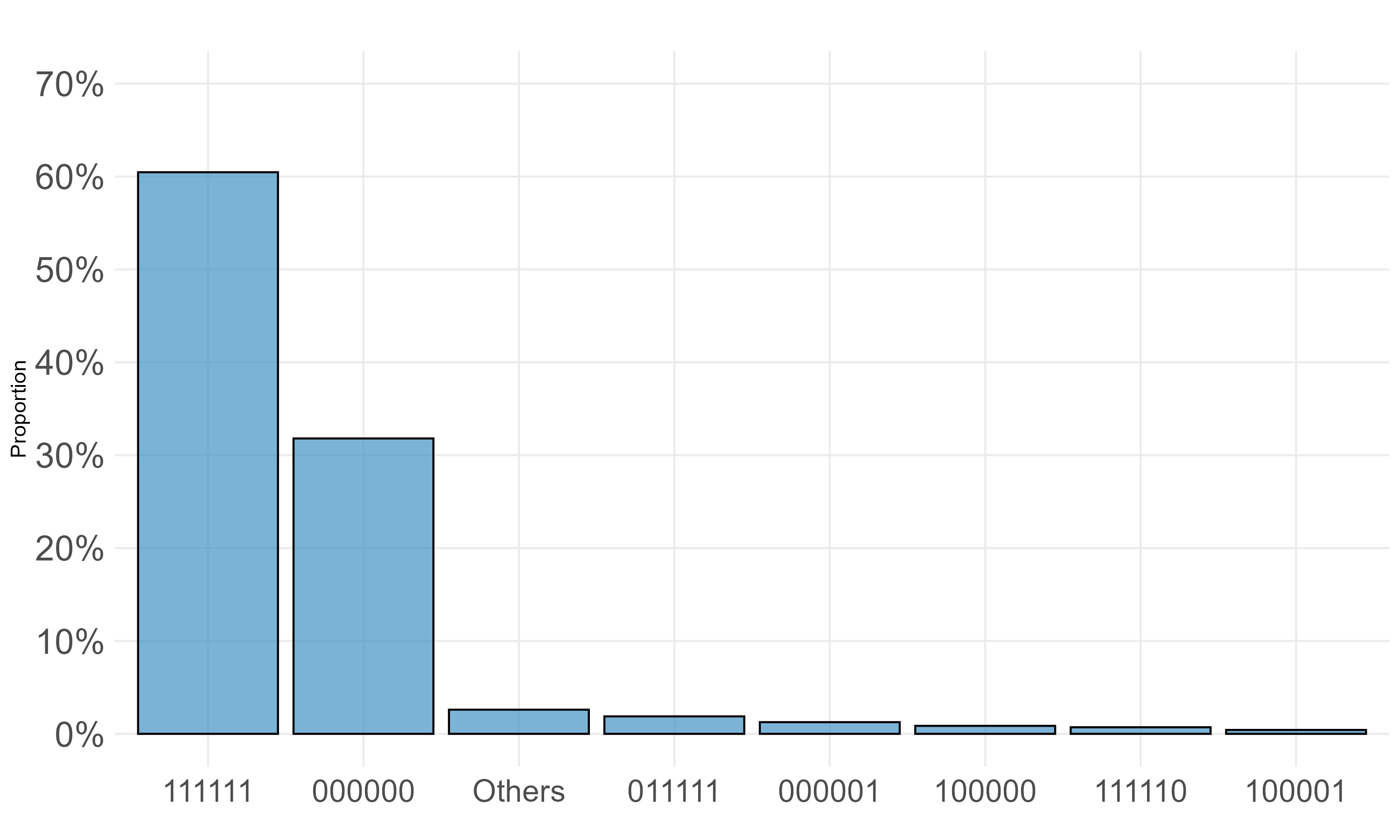}
    \centerline{Panel B: GPT-4o}
    
    \caption{Proportion of Correctness Combinations Across Six Rounds}
    \label{fig:repli_combined}
\end{figure}

For the the pairwise differences across rounds, Table \ref{tab:007_gpt35_rounds_diffference} and Table \ref{tab:007_gpt4o_rounds_diffference} present the difference between two rounds for each model with the average and standard deviation in parenthesis and also p-values.  We also report the detailed distribution of response patterns for these two models in Figure \ref{fig:repli_combined}. Each pattern is represented by a six-digit binary sequence, where '1' indicates a correct answer and '0' indicates an incorrect answer in a given round. For example, '111111' represents consistent correctness across all rounds, while '000000' represents consistent incorrectness.  These pattern distributions reveals high consistency across rounds in binary responses, with GPT-3.5 and GPT-4o demonstrating consistency rates of 87.58\% and 92.3\%, respectively. 

\begin{table}[!h]
\centering
\caption{Differences in Accuracy and Confidence: GPT-3.5}
\centering
\begin{tabular}[t]{lcccc}
\toprule
\multicolumn{1}{c}{ } & \multicolumn{2}{c}{Accuracy} & \multicolumn{2}{c}{Confidence} \\
\cmidrule(l{3pt}r{3pt}){2-3} \cmidrule(l{3pt}r{3pt}){4-5}
Rounds & Mean (SD) & p-value & Mean (SD) & p-value\\
\midrule
1\_2 & 0.001 (0.238) & 0.733 & -0.000 (0.040) & 0.810\\
\addlinespace
1\_3 & 0.002 (0.236) & 0.439 & -0.000 (0.039) & 0.469\\
\addlinespace
1\_4 & 0.002 (0.239) & 0.309 & -0.000 (0.040) & 0.291\\
\addlinespace
1\_5 & 0.002 (0.238) & 0.395 & -0.000 (0.039) & 0.378\\
\addlinespace
1\_6 & -0.002 (0.258) & 0.346 & 0.000 (0.042) & 0.998\\
\addlinespace
2\_3 & 0.001 (0.219) & 0.611 & -0.000 (0.039) & 0.622\\
\addlinespace
2\_4 & 0.002 (0.216) & 0.372 & -0.000 (0.039) & 0.397\\
\addlinespace
2\_5 & 0.001 (0.226) & 0.822 & -0.000 (0.040) & 0.531\\
\addlinespace
2\_6 & -0.003 (0.254) & 0.263 & 0.000 (0.042) & 0.809\\
\addlinespace
3\_4 & 0.001 (0.215) & 0.572 & -0.000 (0.039) & 0.725\\
\addlinespace
3\_5 & -0.001 (0.224) & 0.751 & -0.000 (0.040) & 0.885\\
\addlinespace
3\_6 & -0.005 (0.257) & 0.076 & 0.000 (0.042) & 0.490\\
\addlinespace
4\_5 & -0.001 (0.217) & 0.708 & 0.000 (0.039) & 0.836\\
\addlinespace
4\_6 & -0.005 (0.258) & 0.065 & 0.000 (0.042) & 0.311\\
\addlinespace
5\_6 & -0.004 (0.257) & 0.114 & 0.000 (0.042) & 0.410\\
\addlinespace
\bottomrule
\label{tab:007_gpt35_rounds_diffference}
\end{tabular}
\end{table}

\begin{table}[!h]
\centering
\caption{Differences in Accuracy and Confidence: GPT-4o}
\centering
\begin{tabular}[t]{lcccc}
\toprule
\multicolumn{1}{c}{ } & \multicolumn{2}{c}{Accuracy} & \multicolumn{2}{c}{Confidence} \\
\cmidrule(l{3pt}r{3pt}){2-3} \cmidrule(l{3pt}r{3pt}){4-5}
Rounds & Mean (SD) & p-value & Mean (SD) & p-value\\
\midrule
1\_2 & 0.012 (0.213) & 0.000 & -0.006 (0.037) & 0.000\\
\addlinespace
1\_3 & 0.012 (0.216) & 0.000 & -0.006 (0.037) & 0.000\\
\addlinespace
1\_4 & 0.010 (0.215) & 0.000 & -0.007 (0.036) & 0.000\\
\addlinespace
1\_5 & 0.009 (0.211) & 0.000 & -0.006 (0.039) & 0.000\\
\addlinespace
1\_6 & 0.017 (0.238) & 0.000 & -0.003 (0.038) & 0.000\\
\addlinespace
2\_3 & -0.000 (0.096) & 0.917 & -0.000 (0.029) & 0.760\\
\addlinespace
2\_4 & -0.002 (0.116) & 0.143 & -0.000 (0.031) & 0.183\\
\addlinespace
2\_5 & -0.002 (0.112) & 0.033 & -0.000 (0.030) & 0.796\\
\addlinespace
2\_6 & 0.005 (0.189) & 0.007 & 0.003 (0.035) & 0.000\\
\addlinespace
3\_4 & -0.002 (0.117) & 0.170 & -0.000 (0.028) & 0.259\\
\addlinespace
3\_5 & -0.002 (0.107) & 0.032 & 0.000 (0.029) & 0.975\\
\addlinespace
3\_6 & 0.005 (0.195) & 0.008 & 0.003 (0.036) & 0.000\\
\addlinespace
4\_5 & -0.001 (0.114) & 0.538 & 0.000 (0.030) & 0.273\\
\addlinespace
4\_6 & 0.007 (0.199) & 0.001 & 0.003 (0.037) & 0.000\\
\addlinespace
5\_6 & 0.008 (0.196) & 0.000 & 0.003 (0.037) & 0.000\\
\addlinespace
\bottomrule
\label{tab:007_gpt4o_rounds_diffference}
\end{tabular}
\end{table}

\clearpage
\subsubsection{Regression Analysis of Response Persistence}

To formally assess the stability of LLM responses across 5 rounds, we conduct three types of regression analyses for both GPT-3.5 and GPT-4o. Table \ref{tab: 008_replication_regression} presents the results from these analyses.

Model 1 (columns 1-2) examines the persistence in response accuracy across rounds. The strong coefficients on previous accuracy (0.7675 for GPT-3.5 and 0.8246 for GPT-4o, both significant at the 0.001 level) indicate high consistency in whether responses are correct across rounds. 

Model 2 (columns 3-4) analyzes how previous accuracy influences confidence levels in subsequent rounds. While statistically significant, the small coefficients (0.0102 for GPT-3.5 and 0.0273 for GPT-4o) suggest that previous accuracy has limited impact on future confidence levels. The high constant terms (0.9400 and 0.9230) indicate that both models maintain high baseline confidence levels regardless of previous performance.

Model 3 (columns 5-6) combines both effects by including both previous confidence and previous accuracy on current confidence. The results reveal strong persistence in confidence levels, particularly for GPT-4o (coefficient of 0.7308 compared to 0.5237 for GPT-3.5). Once controlling for previous confidence, the effect of previous accuracy becomes small, suggesting again that confidence patterns are largely independent of actual performance.

Across all specifications, round effects are minimal, with coefficients close to zero, which indicates that the sequence of questions does not systematically influence either accuracy or confidence.
\begin{table}[!h]
\centering
\caption{Response Persistence and Round Effects}
\centering
\begin{tabular}[t]{lcccccl}
\toprule
\multicolumn{1}{c}{ } & \multicolumn{2}{c}{Response Correctness} & \multicolumn{2}{c}{Confidence Level} & \multicolumn{2}{c}{Combined Analysis} \\
\cmidrule(l{3pt}r{3pt}){2-3} \cmidrule(l{3pt}r{3pt}){4-5} \cmidrule(l{3pt}r{3pt}){6-7}
 & GPT-3.5  & GPT-4o  & GPT-3.5 & GPT-4o & GPT-3.5  & GPT-4o \\
\midrule
Previous Accuracy & 0.7675*** & 0.8246*** & 0.0102*** & 0.0273*** & 0.0038*** & 0.0049***\\
 & (0.0036) & (0.0027) & (0.0006) & (0.0010) & (0.0004) & (0.0004)\\
\addlinespace
Previous Confidence &  &  &  &  & 0.5237*** & 0.7308***\\
 &  &  &  &  & (0.0087) & (0.0085)\\
\addlinespace
Round 3 & 0.0009 & 0.0158** & -0.0007 & -0.0063*** & -0.0006 & -0.0043***\\
\addlinespace
 & (0.0052) & (0.0051) & (0.0004) & (0.0004) & (0.0005) & (0.0006)\\
\addlinespace
Round 4 & 0.0012 & 0.0143** & -0.0008 & -0.0066*** & -0.0006 & -0.0046***\\
 & (0.0052) & (0.0051) & (0.0004) & (0.0004) & (0.0005) & (0.0006)\\
\addlinespace
Round 5 & -0.0007 & 0.0148** & -0.0007 & -0.0062*** & -0.0005 & -0.0040***\\
 & (0.0052) & (0.0051) & (0.0004) & (0.0004) & (0.0005) & (0.0006)\\
\addlinespace
\addlinespace
Round 6 & -0.0044 & 0.0231*** & -0.0004 & -0.0033*** & -0.0002 & -0.0013\\
 & (0.0055) & (0.0055) & (0.0004) & (0.0004) & (0.0005) & (0.0007)\\
\addlinespace
Question Order & 0.0000*** & 0.0000*** & -0.0000*** & -0.0000*** & -0.0000* & -0.0000***\\
 & (0.0000) & (0.0000) & (0.0000) & (0.0000) & (0.0000) & (0.0000)\\
\addlinespace
Constant & 0.0628*** & 0.0895*** & 0.9400*** & 0.9230*** & 0.4482*** & 0.2531***\\
\addlinespace
 & (0.0053) & (0.0066) & (0.0006) & (0.0010) & (0.0082) & (0.0082)\\
\addlinespace
\hline
\addlinespace
N & 58215 & 59996 & 58794 & 59967 & 58791 & 59938\\
R$^2$ & 0.593 & 0.681 & 0.012 & 0.055 & 0.280 & 0.553\\
\addlinespace
\bottomrule
\end{tabular}
\label{tab: 008_replication_regression} 
\begin{tablenotes}
\small
\item \textit{Notes:} $^{***}$p $<$ 0.001; $^{**}$p $<$ 0.01; $^{*}$p $<$ 0.05.
\end{tablenotes}
\end{table}

\end{document}